\documentclass[12pt]{article}
\pdfoutput=1
\usepackage[a4paper]{geometry}
 
\usepackage[T1]{fontenc} 
\usepackage{comment}

\usepackage{jheppub,hyperref,float,array,adjustbox,mathtools,physics, xcolor}

\usepackage{xcolor,tikz,pgfplots,amsmath,amsfonts}
\usepackage{graphicx}
\usepackage[mathscr]{euscript}
\usetikzlibrary{matrix,calc,positioning,decorations.markings,decorations.pathmorphing,decorations.pathreplacing}
\usetikzlibrary{arrows,cd}
\usetikzlibrary{shapes.misc}
\usetikzlibrary {shapes.geometric}
\usetikzlibrary {shapes.multipart}
\usetikzlibrary{decorations.markings}
\usetikzlibrary{backgrounds}
\usetikzlibrary{shapes.gates.logic.US}
\usetikzlibrary{circuits.ee.IEC}
\usepackage{bbding}

\definecolor{terradisiena}{RGB}{233,116,81}
\definecolor{strisciadipietro}{RGB}{229,204,255}
\definecolor{verdepetrolio}{RGB}{33,100,119}

\usepackage{tikz}
\usetikzlibrary{automata,positioning,calc}
\usetikzlibrary{decorations.markings}
\tikzset{->-/.style={decoration={markings, mark=at position #1 with {\arrow{>}}},postaction={decorate}}}
\tikzset{-<-/.style={decoration={markings, mark=at position #1 with {\arrow{<}}},postaction={decorate}}}
\tikzset{auto shift/.style={auto=right,->, to path={ let \p1=(\tikztostart),\p2=(\tikztotarget), \n1={atan2(\y2-\y1,\x2-\x1)},\n2={\n1+180} in ($(\tikztostart.{\n1})!1mm!270:(\tikztotarget.{\n2})$) -- ($(\tikztotarget.{\n2})!1mm!90:(\tikztostart.{\n1})$) \tikztonodes}}}

\usetikzlibrary {shapes.geometric}

\newcommand{\SU}[1]{\mathrm{SU}(#1)}
\newcommand{\U}[1]{\mathrm{U}(#1)}

\newcommand{\numberset}{\mathbb}
\newcommand{\N}{\numberset{N}}
\newcommand{\Z}{\numberset{Z}}

\newcommand{\C}{\numberset{C}}

\newcommand{\ee}{\mathrm{e}}
\newcommand{\mi}{\mathrm{i}}
\renewcommand{\Re}{\operatorname{Re}}
\renewcommand{\Im}{\operatorname{Im}}
\newcommand{\fr}[1]{\mathfrak{#1}}

\usepackage{float}

\title{
\begin{center}
Cardy matches Bethe on the Surface:\\
\Large{a Tale of a Brane and a Black Hole}
\end{center}
}

\author[a]{Antonio Amariti,}	
\author[a,b]{Pietro Glorioso,}
\author[a,b]{Davide Morgante}
\author[a,b]{and Andrea Zanetti}

\affiliation[a]{INFN, Sezione di Milano, Via Celoria 16, I-20133 Milano, Italy}
\affiliation[b]{Dipartimento di Fisica, Università degli studi di Milano, Via Celoria 16, I-20133, Milano, Italy}

\emailAdd{antonio.amariti@mi.infn.it}
\emailAdd{pietro.glorioso@mi.infn.it}
\emailAdd{davide.morgante@mi.infn.it}
\emailAdd{andrea.zanetti@mi.infn.it}

\abstract{
   We consider the insertion of a Gukov-Witten surface defect in $\mathrm{SU}(N)$ $\mathcal{N}=4$ SYM corresponding to a probe D3-brane in the holographic dual setup. The defect gives rise to a $4d$-$2d$ coupled system encoding the entropy of the dual perturbed black hole, which can be extracted from the corresponding Superconformal Index. Elaborating on previous studies, we refine the results using both a saddle-point and a Bethe-Ansatz approach. The consistency of our computation is corroborated by the complete agreement between the two results in the appropriate regime of fugacities. Eventually, the sub-leading structure, emerging from our analysis, provides a suggestive EFT interpretation for the addition of the defect to the $4d$ system, mirroring the behavior of the probe D3-brane in the gravity dual. 
}

\begin{document}
\maketitle
\flushbottom
\allowdisplaybreaks 

\section{Introduction}

In the recent past important progress in the microstates counting of $5d$ rotating charged supersymmetric black hole has been possible thanks to the role played by the Superconformal Index (SCI) of \cite{Kinney:2005ej,Romelsberger:2005eg}. Such a proliferation of results spread after the seminal work of \cite{Hosseini:2017mds} where an entropy function, counting microstates of the dual black hole, was proposed. The microscopic origin of the Bekenstein-Hawking entropy of the holographic dual supersymmetric black hole was provided in \cite{Cabo-Bizet:2018ehj}.

Motivated by these results it became crucial to extract the entropy function of \cite{Hosseini:2017mds} from a pure field theory calculation. The SCI, even if expected to be the natural candidate for this computation, initially failed to provide the $\order{N^2}$ scaling of the microscopic degrees of freedom \cite{Kinney:2005ej} due to large cancellations among states with opposite statistics. The resolution of the puzzle was found for $\SU{N}$ $\mathcal{N}=4$ SYM in \cite{Choi:2018hmj, Benini:2018ywd} by using two different methodologies. Based on these, one can then distinguish two broad classes of computations of the index for supersymmetric theories in $4d$. Either one first compute the integral exactly and then evaluate the leading contribution to the entropy \cite{Benini:2018mlo,Benini:2018ywd,Aharony:2024ntg,Mamroud:2022msu,Colombo:2021kbb,Aharony:2021zkr,Benini:2021ano,Lezcano:2021qbj,Lezcano:2019pae,Lanir:2019abx,Benini:2020gjh} or one evaluates the integrand and then extract the entropy from a saddle point analysis \cite{Choi:2023tiq,Cabo-Bizet:2020nkr,Amariti:2023rci,GonzalezLezcano:2022hcf,Ardehali:2021irq,Jejjala:2021hlt,Goldstein:2020yvj,Cabo-Bizet:2019eaf,Cabo-Bizet:2020ewf,Amariti:2019mgp,Kim:2019yrz,ArabiArdehali:2019tdm,Choi:2022asl,Honda:2019cio,GonzalezLezcano:2020yeb,Amariti:2020jyx,Amariti:2021ubd}.
The first approach, originally discussed in \cite{Benini:2018mlo,Benini:2018ywd}, provides, in principle, an exact answer in any regime of charges. However, the formal exact evaluation turns out to be rather complicated and it boils down to solve a set of algebraic equation, referred in the literature as  Bethe Ansatz Equations (BAEs). Despite such difficulties in the case of $\SU{N}$ $\mathcal{N}=4$ SYM the solutions are known at large $N$ and it has allowed to extract the black hole entropy, matching it with the gravitational expectations \cite{Gutowski:2004yv,Gutowski:2004ez}.
A simpler calculation, valid only in a restricted regime of charges, corresponds to the so called Cardy-like limit \cite{Choi:2018hmj}. In this case one estimates the integral from a saddle point analysis and then the entropy can be obtained also at finite $N$. 
Furthermore, a third method consists of a direct saddle point evaluation of the matrix integral at large $N$  in \cite{Cabo-Bizet:2019eaf}. 
Observe that the saddle point evaluations of the index can be generalized to other $4d$ models with different matter content and supersymmetry and such results inspired the EFT calculations of \cite{Cassani:2021fyv,ArabiArdehali:2021nsx} on  the high temperature limit on the second sheet of the index.

Despite the power of the results discussed so far it is desirable to go beyond, by perturbing the black hole, and as a consequence the superconformal index in a controlled way.
For example, perturbing the system with the addition of a Polyakov loop provides an order parameter to detect the confinement/deconfinement transition, expected to correspond to the dual mechanism of the (first order) Hawking-Page transition from the thermal AdS to
the large black hole \cite{Witten:1998zw,Sundborg:1999ue,Aharony:2003sx}
(see \cite{Choi:2018vbz,Chen:2022hbi,Copetti:2020dil,Choi:2021lbk,Perez-Garcia:2024pcq}
for recent progresses in the understanding of the Hawking-Page transition from the field theory side).

Recently another (supersymmetry preserving) order parameter for the deconfinement phase transition has been proposed in \cite{Chen:2023lzq} by adding a surface defect corresponding on the gravitational side to a 
probe D3-brane, extended across the time and a radial direction, and wrapped on one compact direction in AdS$_5$ and one compact direction in $S^5$.
Such a probe D3 is interpreted in the dual field theory as a half BPS Gukov-Witten surface defect  placed on $\mathbb{R}^2$ at $x_2=x_3=0$ in $\mathbb{R}^{1,3}$. The defect corresponds to a codimension-2 singularity in $\mathcal{N}=4$ SYM.
A class of Gukov-Witten defects in $\mathcal{N}=4$ SYM is classified by specifying its Levi subgroup embedding. 
The defect studied in  \cite{Chen:2023lzq}  corresponds to the maximal Levi subgroup embedding.
From the $2d$ SCFT point of view the theory living on the defect is then an $\mathcal{N}=(4,4)$ $\U{1}$ gauge theory with $N$ fundamental hypermultiplets\footnote{See also \cite{Kim:2024ucf} for a similar setup where the actual defect corresponds to the one studied in \cite{Nakayama:2011pa}}. 
The surface operator defined in this way  probes a $1/16$ BPS black hole in the generalized thermal ensemble given by the superconformal index.
At technical level the effect of the surface defect on the index is obtained by coupling the $4d$-$2d$ system along the lines of the discussion of \cite{Gadde:2013dda}.
This coupling is done by gauging the global symmetry of the $2d$ theory, identifying it to the $\SU{N}$  gauge symmetry of  4d  $\mathcal{N}=4$ SYM.
The final expression for the superconformal index of the $4d$-$2d$ coupled system is then given by the original index and in addition to the integrand the contribution of opportune insertions of Jacobi $\theta_0$ functions carrying the charges of the $2d$ fields once expressed in terms of the 4d ones.

From the gravitational side the leading contribution of the probe D3 to the free energy of the Black Hole has been computed in \cite{Chen:2023lzq} in the case of equal charges and different angular momenta. The final result correspond to a sum of the unperturbed result and to the perturbative contribution from the DBI action.  
Translating the result to the entropy in turns put that the charge and the entropy of the D3 are complex in this case. 
The result is apparently contradictory with respect with the one obtained from the field theory side using the superconformal index, where such quantities are real.
The way out of the apparent contradiction has been subsequently discussed in \cite{Cabo-Bizet:2023ejm}, where it has been shown that by  borrwing the field theory 
result, where the entropy is obtaind from the laplace transform of the superconformal index, also the gravitational entropy can be shown to be real.

Furthermore while in \cite{Chen:2023lzq} the evaluation of the defect superconformal  index has been pursued using a  direct saddle point evaluation at $1/N$ order in a fixed regime of charges, in  \cite{Cabo-Bizet:2023ejm} the evaluation of the superconformal index has been done through a systematic Cardy-like expansion with more generic regimes of charges allowed.

While the two result match in the regime of small angular momenta at fixed charges, 
this second approach is intriguing because it  tells us more informations about the backreaction of the probe D3,
predicting a fully backreacted answer at leading order in the Cardy like limit.
The result furthermore suggests the structure of the subleading correction in the Cardy-like limit in terms of the angular fugacities.
Namely a series expansion in the angular momenta could be derived, 
going beyond the leading order approximation.

In this paper we discuss such backreaction effect from the SCI perspective, evaluating the SCI in the Cardy-like limit and with the Bethe Ansatz (BA) approach. To this end we compute such backreaction in the Cardy-like limit around the holonomy saddles giving rise to the black hole. The result is compatible with \cite{Cabo-Bizet:2023ejm} and it allows to estimate the $3d$ partition function with the addition of the effect of the defect. Surprisingly we find that the two regions corresponding to the second and the third sheet from the EFT approach give rise to the same result, symmetrizing the seeming asymmetry of \cite{Cabo-Bizet:2023ejm}.\\
Then we confirm our result following the Bethe Ansatz approach.
We first verify its feasibility in presence of the defect and then we compute the contribution of the basic solution of the BAEs. 
The computation is done for equal, not necessarily large, angular momenta and for arbitrary flavor charges.
Thus, we generalize the result obtained for equal and fixed flavor charges in \cite{Chen:2023lzq}.
The resulting index, once evaluated on the basic solution, recovers its symmetry in agreement with the Cardy-like analysis.

\section{Gukov-Witten Surface Operators}
\label{sec:davide}
\subsection{Field Theory Construction}
In this section we give a brief review of the construction of two-dimensional operators by Gukov and Witten (GW) \cite{Gukov:2008sn,Gukov:2006jk}. We then discuss their field theory interpretation as a coupled $4d$-$2d$ system and the computation of the SCI in the presence of such defects \cite{Gadde:2013dda}.

Extended operators in a QFT can be broadly divided into two sub-categories: either defined by functionals of local operators on some higher co-dimension manifold or as singularities in the gauge field. From this classification one recognizes Wilson and 't Hooft line operators of four-dimensional pure Yang-Mills theory. The former are classified by representations of the gauge group $G$, while the latter are, generically, classified by integers labelling the amount of magnetic charge. \\
Gukov and Witten gave a prescription to generalize this construction to two dimensional surface operators as singularities for the vector field on a surface $\Sigma$. Surface defects, in contrast to line defects, are classified not only by the singularity of the vector field along $\Sigma$, but also by the subgroup of $G$ under which they are invariant. In this section we will consider such defects in $4d$ $\mathcal{N}=4$ SYM even if the construction is more general.\\
We regard the field content of $\mathcal{N}=4$ SYM in $\mathcal{N}=1$ language where we the field content is given by a vector multiplet $V$ and chiral multiplets $\Phi_{i=1,2,3}$. A half-BPS GW surface operator oriented along the $(x^0,x^1)$ direction, is defined as a singularity on the vector field and the scalar component of the chiral multiplets
\begin{equation}
    A=a(r)\dd{\theta}+\cdots,\qquad \phi=b(r)\frac{\dd{r}}{r}-c(r)\dd{\theta}+\cdots\; ,
    \label{eq:Aphising}
\end{equation}
where $A=A_\mu\dd{x}^\mu,\;\phi=\phi_\mu\dd{x}^\mu$ with $\mu=2,3$ and $z\equiv r\ee^{\mi \theta}=x^2+\mi x^3$ is the normal direction to $\Sigma$. The BPS condition can be casted in the form of Hitchin equations
\begin{equation}
    \begin{cases}
        F_A-\phi\wedge\phi=0,\\
        \dd_A\phi=0,\quad \dd_A\star\phi=0
    \end{cases}
    \label{eq:BPS4d}
\end{equation}
and conformal invariance requires $a,b,c$ in \eqref{eq:Aphising} to be independent on $r$. Hitchin equations also require $a,b,c$ to be mutually commuting. The easiest solution to \eqref{eq:Aphising} is obtained by conjugating the algebra-valued parameters $a,b,c$ to parameters $\alpha,\beta,\gamma$ valued in the Lie algebra $\fr{t}$ of the maximal torus $\mathbb{T}$ of the gauge group $G$. Therefore, the singularity is described by 
\begin{equation}
    A=\alpha\dd{\theta}+\cdots,\qquad \phi =\beta \frac{\dd{r}}{r}-\gamma \dd{\theta}+\cdots\;.
    \label{eq:singularity}
\end{equation}
Furthermore, it turns out that one can add a $2d$ $\theta$-term labelling topologically distinct restrictions of the $G$-bundle to the defect. The parameter $\eta$ labelling this choice generically takes value in a subgroup of $^L\mathbb{T}$, the maximal torus of the Langlands dual of $G$. This fully defines the insertion of a GW surface defect in the path integral. However, when summing over gauge configurations in the path integral, one divides by the subgroup of $G$ commuting with the parameters $\alpha,\beta,\gamma,\eta$. This condition defines the subgroup that preserves the singularity which is denoted as Levi subgroup $\mathbb{L}$. The choice of $\mathbb{L}$ is regarded as the definition of the defect. Therefore, the insertion of the defect amounts to a choice of 
\begin{equation}
    (\alpha,\beta,\gamma,\eta)\in (\mathbb{T}\times\fr{t}\times\fr{t}\times {^L}\mathbb{T})/\text{Weyl}(\mathbb{L}).
\end{equation}

In the rest of the paper we focus on $G=\SU{N}$. The Levi subgroups in this case are classified by partitions of $N=\lambda_1+\cdots+\lambda_s$. A partition $\lambda=\qty[\lambda_1,\ldots,\lambda_s]$, with $0<\lambda_1\le \lambda_2\le\cdots\le \lambda_n<N$, is associated to the Levi subgroup
\begin{equation}
    \mathbb{L}=\mathrm{S}\qty(\bigotimes_{i=1}^n \U{k_i}),\qquad N=\sum_{i=1}^n k_i\,,
    \label{eq:leviSU}
\end{equation}
where $k_i$ are the number of boxes in the $i$-th column Young tableaux associated to $\lambda$. For example, let us consider $G=\SU{5}$ and $\lambda=\qty[4,1]$. Here the Young tableaux is 
\begin{equation}
   \begin{tikzpicture}[scale=.7]
        \foreach \i in {0,1,2,3,4}
            \draw[thick] (\i,0) rectangle (1,1);
        \draw[thick] (0,-1) rectangle (1,1);

        \node[left] at (0,.5) {$4$};
        \node[left] at (0,-.5) {$1$};
        \node[above] at (.5,1) {$2$};
        \node[above] at (1.5,1) {$1$};
        \node[above] at (2.5,1) {$1$};
        \node[above] at (3.5,1) {$1$};
   \end{tikzpicture} 
\end{equation}
and the associated Levi subgroup is 
\begin{equation}
    \mathbb{L}=\mathrm{S}\qty(\U{2}\times \U{1}\times\U{1}\times\U{1})\simeq \SU{2}\times\U{1}^3.
\end{equation}
For $\SU{N}$, the Levi subgroup associated to the partition $\lambda=\qty[N-1,1]$ is dubbed "maximal Levi sub-group". This is going to be the relevant sub-group for our analysis.

The $2d$ defect is coupled to the $4d$ theory by imposing the singular behavior on the $4d$ gauge fields in the path integral. In practice this is quite cumbersome and usually one can add a $2d$ theory acting as a Lagrange multiplier such to impose the singular behavior on $\Sigma$. This $2d$ gauge theory is built so that, when integrating out the excitation on the defect, one recovers the original $4d$ theory with the constrained fields. \\
The $2d$ theory must satisfy certain properties \cite{Gadde:2013dda,Gukov:2008sn,Gukov:2006jk} to prescribe the singularity as described above. Let us consider the scenario where the $2d$ theory is a Gauge Linear Sigma Model (GLSM) with some target space $\mathscr{M}_{\alpha,\beta,\gamma}$. A half-BPS defect must preserve $\mathcal{N}=(4,4)$ supersymmetry, implying that the target space of the GLSM must be hyper-Kähler. Additionally, it must possess a $G$ action, suggesting that the $2d$ theory must have a $G$ flavor symmetry that is used to couple it to the $4d$ bulk. Furthermore, it must be dependent on the choice of $\mathbb{L}$. The simplest target space is the cotangent space of $G/\mathbb{L}$, denoted as $T^*(G/\mathbb{L})$. It is worth noting that $T^*(G/\mathbb{L})=G_{\mathbb{C}}/\mathbb{L}_{\mathbb{C}}$ is also as the moduli space of solutions with the prescribed singularity of the form \eqref{eq:singularity}. The action of the $2d$ system can be obtained straightforwardly: the coupling to the $2d$ theory induces a singularity in the BPS equations \eqref{eq:BPS4d} 
\begin{equation}
    F_{23}+\comm{\phi_2}{\phi_2^\dagger}=2\pi \delta^{(2)}(\vec{x})q q^\dagger, \qquad D_{\bar{z}}\phi_2=\pi \delta^{(2)}(\vec{x})q\tilde{q}\,,
\end{equation}
where $q q^\dagger$ and $q\tilde{q}$ are moment maps for the $G$ action on $\mathscr{M}_{\alpha,\beta,\gamma}$. By virtue of the BPS equations of the $2d$ theory, the moment maps are integrated out in favour of the Kähler moduli $\alpha+\mi \eta$ and $\beta+\mi \gamma$ respectively. This, together with the $\delta$-functions, induces the singular behavior of the solution \eqref{eq:singularity}.

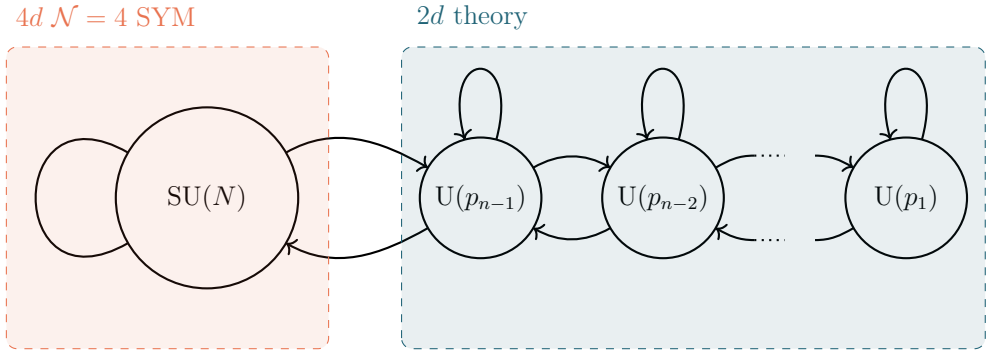
\begin{figure}
    \centering
    \begin{tikzpicture}[scale=.8,every node/.style={scale=0.85}]

        \draw[thick] (.5,0) circle (1.5);
        \node at (.5,0) {$\SU{N}$};
        \draw[thick, -] ($(.5,0)+(150:1.5)$) to[out=150, in=210, looseness=4] ($(.5,0)+(210:1.5)$);

        \foreach \x/\lab in {5/p_{n-1},8/p_{n-2},12/p_1}{
            \draw[thick] (\x,0) circle (1);
            \node at (\x,0) {$\U{\lab}$};
            \draw[thick, ->] ($(\x,0)+(75:1)$) to[out=75, in=105, looseness=8] ($(\x,0)+(105:1)$);}
        
        \draw[thick, ->] ($(.5,0)+(30:1.5)$) to[out=30,in=150] ($(5,0)+(150:1)$);
        \draw[thick, <-] ($(.5,0)+(-30:1.5)$) to[out=-30,in=-150] ($(5,0)+(-150:1)$);

        \draw[thick, ->]  ($(5,0)+(30:1)$) to[out=30,in=150] ($(8,0)+(150:1)$);
        \draw[thick, <-]  ($(5,0)+(-30:1)$) to[out=-30,in=-150] ($(8,0)+(-150:1)$);

        \draw[thick, -]  ($(8,0)+(30:1)$) to[out=30,in=180] ($(9.5,0)+(90:.7)$);
        \draw[thick, <-]  ($(8,0)+(-30:1)$) to[out=-30,in=-180] ($(9.5,0)+(-90:.7)$);
        \draw[thick, -, dotted]  ($(9.5,0)+(90:.7)$) to[out=0,in=180] ($(10,0)+(90:.7)$);
        \draw[thick, -, dotted]  ($(9.5,0)+(-90:.7)$) to[out=0,in=180] ($(10,0)+(-90:.7)$);

        \draw[thick, <-]  ($(12,0)+(150:1)$) to[out=150,in=0] ($(10.5,0)+(90:.7)$);
        \draw[thick, -]  ($(12,0)+(-150:1)$) to[out=-150,in=0] ($(10.5,0)+(-90:.7)$);

        \node[right, terradisiena] at (-2.8, 3) {$4d$ $\mathcal{N}=4$ SYM};
        \node[right, verdepetrolio] at (3.8, 3) {$2d$ theory};

        \draw[fill=terradisiena, fill opacity=.1, terradisiena, dashed, rounded corners] (.5,0)+(-3.3,-2.5) rectangle +(2,2.5);
        \draw[fill=verdepetrolio, fill opacity=.1, verdepetrolio, dashed, rounded corners] (8.5,0)+(-4.8,-2.5) rectangle +(4.8,2.5);
    \end{tikzpicture}
    \caption{Quiver diagram of the $4d$-$2d$ coupled system in the $\mathcal{N}=(2,2)$ language in $2d$ and $\mathcal{N}=2$ language in $4d$. Notice that the ranks of the gauge groups of the $2d$ theory go from right to left.}
    \label{fig:quiver}
\end{figure}

In order to describe the  gauge group $G_{2d}$ of the $2d$ theory, one further needs to describe $\mathscr{M}_{\alpha,\beta,\gamma}$ as an hyper-Kähler quotient of some vector space by the group $G_{2d}$. For the case at hand, where the $4d$ gauge theory is $\mathcal{N}=4$ $\SU{N}$ SYM, this quotient can always be constructed \cite{Kobak:1996} and the resulting gauge theory is a $2d$ $\mathcal{N}=(4,4)$ theory with flavor symmetry $G=\SU{N}$ and gauge group 
\begin{equation}
    G_{2d}=\bigotimes_{i=1}^{n-1}\U{p_i},\quad\text{where}\quad p_i=\sum_{j=1}^i k_j.
\end{equation}
The matter content is given by bi-fundamental hypermultiplets in the $(\mathbf{p}_i,\mathbf{p}_{i+1})$ representation and $N$ fundamental hypermultiplets for $\U{p_{n-1}}$. In $\mathcal{N}=(2,2)$ language the theory is given by the quiver diagram in figure \ref{fig:quiver}. The theory also carries a Weyl anomaly \cite{Wang:2020xkc,Chalabi:2020iie} $c_{2d}$ which for a defect of the form \eqref{eq:leviSU} is given by 
\begin{equation}
    c_{2d}=3\qty(N^2-\sum_{i=1}^n k_i^2).
    \label{eq:centralcharge}
\end{equation}

The construction just described is necessary in order to define the SCI of the $4d$-$2d$ coupled system.
\subsection{\texorpdfstring{The $2d$ Elliptic Genus}{Elliptic}}
Here we briefly discuss how the SCI computation changes in the presence of the GW defect. Firstly, to preserve the right supercharges on $S^3\times S^1$ the GW defect has to be wrapped appropriately on this geometry. This corresponds to wrapping once the defect around the thermal circle and once around the great circle of $S^3$. By the state-operator correspondence, the Hilbert space of the theory is now twisted by the presence of the defect. Therefore, the SCI must be computed on the Hilbert space $\mathcal{H}_\text{GW}(S^3)$. However, this approach is quite involved and one usually exploits the fact that the index of the $4d$-$2d$ coupled system can be casted in the following form
\begin{equation}
    \mathcal{I}(p,q,v_i)=\int_{\SU{N}}\dd{u}\mathcal{I}_{4d}(p,q,v_i;u)\mathcal{I}_{2d}(p,q,v_i;u)\,,
    \label{eq:4d2dindex}
\end{equation}
where $\mathcal{I}_{2d}$ is the index of the $2d$ theory living on the defect, which is wrapped along a temporal $T^2$ in $S^1\times S^2$. 

To fully compute the SCI \eqref{eq:4d2dindex} one needs to understand how the superconformal algebra of the $2d$ $\mathcal{N}=(4,4)$ theory is embedded in the one of $4d$ $\mathcal{N}=4$. The insertion of an half-BPS GW operator in $4d$ breaks the superconformal algebra in 
\begin{equation}
    \fr{u}(1)_A\ltimes(\fr{psu}(1,1|2)\times\fr{psu}(1,1|2))\ltimes\fr{u}(1)_C\subset\fr{psu}(2,2|4)
    \label{eq:embedding}
\end{equation}
which is the usual $2d$ $\mathcal{N}=(4,4)$ superVirasoro algebra, centrally extended by $\fr{u}(1)_C$. The abelian factor $\fr{u}(1)_A$ acts as the outer-automorphism of the algebra and need not be a realized symmetry on the defect. 

The $2d$ contribution to \eqref{eq:4d2dindex} can be found by the following $2d$ index
\begin{equation}
    \mathcal{I}_\text{NSNS}=\Tr_\text{NSNS}(-1)^F\ee^{2\pi \mi \tau_{2d}L_0}\ee^{-2\pi \mi \bar\tau_{2d}(\bar L_0-\frac{1}{2}\bar J_0)}\ee^{2\pi \mi z_{\text{NS}}J_0}\ee^{2\pi \mi \chi J_A}\ee^{2\pi \mi u C}\,,
    \label{eq:NSNSindex}
\end{equation} 
where $\tau_{2d}$ is the complex structure of the temporal $T^2$ and the trace is taken over the NSNS-sector. The various bosonic generators of the embedding \eqref{eq:embedding} that enter in the definition of this index are the right- and left-moving Hamiltonians $\bar L_0,L_0$, the Cartan of $\fr{su}(2)_R$ $J_0$ and the flavor symmetry $J_A$ of the GLSM. To ease the computation of such index, we will use the fact that the NSNS-sector and RR-sector indices are related, and that the latter is the elliptic genus of the GLSM. In fact, as discussed in Appendix C of \cite{Chen:2023lzq}, the following relation holds 
\begin{equation}
    \mathcal{I}_\text{NSNS}(\tau_{2d},z_\text{NS},\chi,u)=\ee^{2\pi\mi\tau_{2d}\qty(-\frac{c_{2d}}{24})}\ee^{-\frac{\pi\mi c_{2d}}{3}\qty(z_\text{NS}-\frac{\tau_{2d}}{2})}\mathcal{I}_\text{RR}(\tau_{2d},z_\text{NS}-\frac{\tau_{2d}}{2},\chi,u)\,,
\end{equation}
where $\mathcal{I}_{RR}$ is the RR-sector index (or fully-refined elliptic genus) and one identifies $z_\text{R}\equiv z_\text{NS}-\frac{\tau_{2d}}{2}$. The $2d$ contribution in \eqref{eq:4d2dindex} is then given by 
\begin{equation}
   \mathcal{I}_{2d}(\tau_{2d},z_\text{NS},\chi,u)=\ee^{-\frac{\pi\mi c_{2d}}{3}z_R}\mathcal{I}_\text{RR}(\tau_{2d},z_\text{R},\chi,u).
   \label{eq:phase}
\end{equation}

Following the discussion of \cite{Benini:2013xpa, Chen:2023lzq, Gadde:2013dda}, the elliptic genus of a general $\mathcal{N}=(2,2)$ gauge theory is given by 
\begin{equation}
    Z_{T^2}=\frac{1}{\abs{W}}\sum_{u_*\in\mathscr{M}_\text{sing}^*}\underset{u=u_*}{\operatorname{JK-Res}}\qty(\mathsf{Q}(u_*),\eta)Z_\text{1-loop}\,,
\end{equation}
where the residues are evaluated following the Jeffrey-Kirwan prescription and $\mathsf{Q}(u_*)$ are the chemical potentials for the fields constrained by the poles at $u_*$. The one-loop contributions for the multiplets are
\begin{equation}
    \begin{split}
        Z_{\mathcal{R}}^\text{chiral}(\tau,\zeta, u)&=\prod_{\rho\in\mathcal{R}}\frac{\theta_1(y^{\frac{R}{2}-1}x^\rho;q)}{\theta_1(y^{\frac{R}{2}}x^\rho;q)},\\[5pt]
        Z_{G}^\text{Vector}(\tau,\zeta,u)&=\qty(\frac{2\pi \eta(q)^3}{\theta_1(y^{-1};q)})^{\rank{G}}\prod_{\alpha\in G}\frac{\theta_1(x^\alpha;q)}{\theta_1(y^{-1}x^\alpha;q)}\prod_{a=1}^{\rank{G}}\dd{\mu}_a,\\[5pt]
        Z^\text{Twisted}(\tau,\zeta)&=\frac{\theta_1(y^{-\frac{R_A}{2}+1};q)}{\theta_1(y^{-\frac{R_A}{2}};q)},
    \end{split}  
\end{equation}
where $\eta(q)$ is the Dedekind eta function and $\theta_1(z;q)$ is the Jacobi theta function. These functions depend on fugacities related to the gauge and global symmetries, which are defined by $q=\ee^{2\pi \mi \tau_{2d}}$, $y=\ee^{2\pi\mi \zeta}$, $x_a=\ee^{2\pi\mi \mu_a}$ and $x^\rho=\ee^{2\pi \mi \rho(\mu)}$.\\
For the case of interest, the $2d$ theory associated to the maximal $\SU{N}$ GW operator $\lambda=[N-1,1]$ is a $\U{1}$ gauge theory with a $\mathcal{N}=(4,4)$ hypermultiplet which is just a pair of $\mathcal{N}=(2,2)$ chiral multiplets in the bi-fundamental of $\SU{N}$. Moreover, we have a $\mathcal{N}=(4,4)$ $\U{1}$ vector multiplet, corresponding in the $\mathcal{N}=(2,2)$ language to a chiral multiplet and a vector multiplet\footnote{In the abelian case, this is equivalent to a twisted chiral multiplet.}. 
The one-loop determinant is given by
\begin{equation}
    \begin{split}
        Z_{\text{1-loop}}&=\frac{2\pi\eta(q)^3}{\theta_1(-\zeta;\tau)}\frac{\theta_1(-2\chi;\tau)}{\theta_1(2\chi-\zeta;\tau)}\prod_{i=1}^N\frac{\theta_1(\mu-u_i+\chi-\zeta;\tau)}{\theta_1(\mu-u_i+\chi;\tau)}\frac{\theta_1(-\mu+u_i+\chi-\zeta;\tau)}{\theta_1(-\mu+u_i+\chi;\tau)}\dd{\mu},
    \end{split}
\end{equation}
where $\chi$ is a fugacity associated to the $\U{1}_A$ symmetry in \eqref{eq:embedding}. Here the arguments of the Jacobi theta function are the chemical potentials rather than the fugacities to avoid clutter. The elliptic genus is computed by integrating over the gauge holonomy $\mu$. Computing the residue around $\mu=u_j-\chi$ one gets cancellations between the vector multiplet determinant and fundamentals with $\beta=\alpha$ ending with the following result 
\begin{equation}
    Z_{T^2}(\tau,\zeta,\chi)=\sum_{i=1}^N\prod_{j\neq i}\frac{\theta_1(u_{i j}+\zeta-2\chi;\tau)\theta_1(u_{i j}-\zeta;\tau)}{\theta_1(u_{i j};\tau)\theta_1(u_{i j}-2\chi;\tau)},\quad u_{i j}=u_i-u_j.
    \label{eq:ellipticgenus}
\end{equation}
The last step is to use this result in \eqref{eq:phase} to get the $2d$ contribution. The phase factor in \eqref{eq:phase} depends on the $2d$ central charge, which for the maximal defect is given by $c_{2d}=6(N-1)$ by using formula \eqref{eq:centralcharge}. Here, the phase can be reabsorbed into the Jacobi $\theta_1$-function, using \eqref{eq:theta01}, to get 
\begin{equation}
    \mathcal{I}_{2d}=\sum_{i=1}^N\prod_{j\neq i}^N\frac{\theta_0(u_{ij}+\zeta-2\chi;\tau)\theta_0(u_{ij}-\zeta;\tau)}{\theta_0(u_{ij};\tau)\theta_0(u_{ij}-2\chi;\tau)},\qquad \zeta\equiv z_\text{NS}.
    \label{eq:index2dfinal}
\end{equation}


\section{Cardy-like approach}

\subsection{\texorpdfstring{The $4d$ Superconformal Index}{SCI}}
In this section we compute the Cardy-like limit of the index of the $4d$-$2d$ coupled system, describing the insertion of a GW defect in $\mathcal{N} = 4$ $\SU{N}$ SYM.

The SCI, originally constructed in \cite{Kinney:2005ej} for $4d$ $\mathcal{N}=4$ SYM, can be defined for a generic $4d$ $\mathcal{N} = 1$ SCFT \cite{Romelsberger:2005eg}, choosing one supercharge $\mathcal{Q}$, as refined Witten index of the theory in radial quantization. The index counts the difference between bosonic and fermionic states annihilated by $\mathcal{Q}$ in the Hilbert space of the theory on $S^3$. Explicitly, the SCI, in the notation of Dolan and Osborn \cite{Dolan:2008qi}, is
\begin{equation}
    \label{eq:ind}
    \mathcal{I}_{4d} = \Trace (-1)^F \ee^{-\beta \{\mathcal{Q},\bar{\mathcal{Q}}\}} p^{J_1}q^{J_2}(pq)^{R/2} \prod_{k} v_k^{Q_k},
\end{equation}
where the refinement is obtained by including charges in the commutant of $\{\mathcal{Q},\bar{\mathcal{Q}}\}$. In (\ref{eq:ind}) $J_1$ and $J_2$ correspond to the angular momenta of the $S^3$, $R$ is the $\U{1}$ R-charge and the fugacities $v_k$ parametrize the Cartan subalgebra, with charges $Q_k$, of other generic symmetries the theory may have.

We are interested in the case of $\mathcal{N} = 4$ SYM with $\SU{N}$ gauge group, for which the index (\ref{eq:ind}) takes the form
\begin{equation}
    \mathcal{I}_{4d} = \Trace_{\text{gauge}} (-1)^F \ee^{-\beta \{\mathcal{Q},\bar{\mathcal{Q}}\}} p^{J_1}q^{J_2}(pq)^{R/2} v_1 ^ {Q_1}v_2 ^ {Q_2},
\end{equation}
where the trace is taken over gauge singlets and $Q_1,\, Q_2$ parametrize the Cartan of the $\fr{su}(3) \subset \fr{su}(4)_R$ commuting with $\{\mathcal{Q},\bar{\mathcal{Q}}\}$. In order to discuss the Cardy-like limit of the index we define
\begin{equation}
    p = \ee ^ {2\pi \mi \tau}, \quad  q = \ee ^ {2\pi \mi \sigma}, \quad  v = \ee ^ {2\pi \mi \xi}
\end{equation}
and introduce the chemical potentials $\Delta_a$ associated to the matter fields, given by
\begin{equation}
    \Delta_a = \rho_a(\xi) + \frac{\tau + \sigma}{2}R_a,
\end{equation}
together with the constraint
\begin{equation}
    \label{eq:del_cons}
    \sum_{a = 1}^3 \Delta_a = \sigma + \tau \mod 1.
\end{equation}
The trace over gauge-invariant states can be achieved by integrating over the holonomies of the gauge group. Defining the elliptic gamma functions 
\begin{equation}
    \Gamma(z;p,q) \coloneqq 
    \prod_{m=0}^{\infty}\prod_{n=0}^{\infty}
    \frac{ 1 - p^{m+1}q^{n+1}/z }{ 1 - p^{m}q^{n}z } \, ,
    \qquad 
    \widetilde{\Gamma}(u) \coloneqq \Gamma(\ee^{2 \pi \mi u}; \ee^{2 \pi \mi \tau}, \ee^{2 \pi \mi \sigma})
    \label{gamma}
\end{equation}
and the $q$-Pochhammer symbol
\begin{equation}
    (z;q)_\infty \coloneqq \prod_{k = 0}^{\infty}\left(1 - z q^k\right),
    \label{poch}
\end{equation}
the index can be expressed as an elliptic hypergeometric integral \cite{Dolan:2008qi}
\begin{equation}
    \label{eq:ind_gauge}
    \mathcal{I}(\Delta,\tau,\sigma) = \frac{(p;p)_\infty^{N-1} (q;q)_\infty^{N-1}}{N!} \prod_{a=1}^{3} \widetilde\Gamma(\Delta_a)^{N-1}\int \prod_{i = 1}^{N-1} \dd u_i \frac{\prod_{a=1}^{3}\prod_{i \neq j}^{N}\widetilde\Gamma (u_{ij} + \Delta_a) }{\prod_{i\neq j}^{N}\widetilde\Gamma(u_{ij})},
\end{equation}
where $u_{ij}\coloneqq u_i-u_j$. The integral is taken over the gauge holonomies, subjected to the $\SU{N}$ constraint
\begin{equation}
    \sum_{i=1}^{N} u_i \in \mathbb{Z}.
    \label{eq:suck}
\end{equation}
Expression (\ref{eq:ind_gauge}), although originally defined for purely imaginary modular parameters $\tau$ and $\sigma$, can be extended to the upper-half complex plane $\tau,\sigma \in \mathbb{H}$, and complex $\Delta$. 

Analytic continuation of chemical potentials introduces phases in \eqref{eq:ind}, which in principle allow for obstructions to Bose/Fermi cancellations. The Cardy-like limit is defined as a generalization of the standard hyperbolic limit of the index \cite{Rains:2006dfy,Van_Diejen_2005}, or high-temperature Cardy limit, where an appropriate scaling behavior is assigned to the analytically continued chemical potentials of the theory, so to preserve the aforementioned cancellations \cite{Choi:2018hmj,ArabiArdehali:2019tdm}. Typically, one defines a complexified inverse temperature parameter $\beta$, such that
\begin{equation}
    \tau = \frac{\mi\beta b}{2\pi}, \quad \sigma = \frac{\mi\beta b^{-1}}{2\pi},
\end{equation}
where $b$ can be identified with the squashing parameter of the $S^3$ in the underlying $S^1\times S^3$ geometry.
While usually $b$ is a real positive number, also complex values are allowed. The non-collinearity of $\tau$ and $\sigma$ in this last case is understood as a twisting of the $S^3$ on the $S^1$ as discussed in \cite{Closset:2013vra, Copetti:2020dil, Cassani:2021fyv}. Then, the Cardy-like limit is defined first by choosing a scaling behavior for the flavor chemical potentials. Such scaling is constituted of a constant fixed term, crucial for preventing Bose/Fermi cancellations, and a linear part in $\beta$. Then, the limit sends the complexified inverse temperature $\beta$ to zero. Within this framework the index (\ref{eq:ind}) can be interpreted as the supersymmetric partition function of the theory\footnote{The difference between the two corresponds to a supersymmetric Casimir energy term \cite{Genolini:2016sxe,Assel:2014paa,Assel:2015nca}. Being proportional to the radius of the $S^1$, such contribution can be neglected in a Cardy-like limit evaluation, where a small radius limit is taken from a geometric perspective, dimensionally reducing the theory on the three-sphere.} on $S^1 \times S^3$ background, with appropriate twisted boundary conditions for the fields, corresponding to the fugacities refinements. We make the connection with the underlying geometry explicit by parametrizing the fugacities as
\begin{equation}
    \label{eq:tau_sigma}
    \tau \coloneqq r\omega_1, \quad \quad \sigma \coloneqq r\omega_2,\quad  \quad r \in \mathbb{R},\; \omega_1, \omega_2 \in \mathbb{H}.
\end{equation}
The definition (\ref{eq:tau_sigma}) explicitly singles out the $r$ parameter, identified with the radius of the $S^1$ in the $S^1\times S^3$ background on which the theory is defined, and the  $\omega_1$, $\omega_2$ squashing parameters for the (possibly squashed) $S^3$. Then, the Cardy-like limit is understood geometrically as a dimensional reduction of the Euclidean theory along the thermal circle.

\subsection{\texorpdfstring{The Cardy-like Limit of the $4d$ Index}{Cardy}}
In order to compute Cardy-like limit of the $4d$ $\mathcal{N} = 4$ SYM index we parametrize the flavor chemical potentials as 
\begin{equation}
    \Delta_a = \tilde \Delta_a + r (\hat{\Delta}_a\omega_1 + \check{\Delta}_a\omega_2 )\equiv \tilde \Delta_a + r \bar{\Delta}_a \quad \quad \tilde{\Delta}_a \in \mathbb{R}/\mathbb{Z},\, \hat{\Delta}_a,\, \check{\Delta}_a \in \mathbb{R}.
    \label{eq:chemicalparam}
\end{equation}
Then, the index can be expanded asymptotically as $r$ is sent to zero and expression \eqref{eq:ind_gauge} can be evaluated through a saddle point approach at fixed $N$. In doing so, the large $N$ limit behavior can be easily inferred extracting a dominant saddle configuration, associated to the black hole solution in the dual gravitational theory \cite{Choi:2018hmj}.

In order to evaluate (\ref{eq:ind_gauge}) in the Cardy-like limit, we rewrite the index in terms of an effective action for the matrix model

\begin{equation}
    \mathcal{I} = \frac{1}{N!}\int \prod_{i=1}^{N-1}\dd u_i \, \ee^{S_{4d}(u,\Delta,\tau,\sigma)},
\end{equation}
where 
\begin{eqnarray}
    \label{eq:s_eff}
    S_{4d}(u,\Delta,\tau,\sigma) =&& (N-1)\sum_{a = 1}^{3}\log\widetilde\Gamma(\Delta_a) \,+\, \sum_{a = 1}^{3}\sum_{i \neq j}^{N}\log\widetilde\Gamma(u_{ij} + \Delta_a) \,+\,\nonumber \\
    +&&\sum_{i < j}^{N}\log\theta_0(u_{ij}) + \sum_{i < j}^{N}\log\theta_0(-u_{ij}) + \nonumber \\
    +&&(N-1)\bigg(\log(p;p)_\infty +\, \log(q;q)_\infty\bigg),
\end{eqnarray}
where the special function $\theta_0$ is given by
\begin{equation}
    \label{theta}
    \theta_0(u;\omega) \coloneqq 
    (\ee^{2 \pi \mi u};\ee^{2 \pi \mi \omega})_{\infty} \,
    (\ee^{2 \pi \mi \omega} \ee^{-2 \pi \mi u};\ee^{2 \pi \mi \omega})_{\infty}\,.
\end{equation}
and we employed \eqref{pmad} in appendix \ref{sec:ell_fun}.
The index reduces to a sum of contributions from multiple saddles
\begin{equation}
    \label{eq:cardyexp}
    \mathcal{I} = \sum_{u^*} I_{\mathrm{lead}}(u^*,\Delta,\tau,\sigma) Z_{\mathrm{sub}}, \quad Z_{\mathrm{sub}}\coloneqq \frac{1}{N!}\int \prod_{i=1}^{N-1}\dd \delta u_i I_{\mathrm{sub}}(\delta u)
\end{equation}
in which we can isolate a leading part, at $\mathcal{O}(r^{-2}$), dependent only on $\Delta$ and the details of the holonomy saddle $u^*$, and a subleading term, $\sim \mathcal{O}(r^0)$, $I_{\mathrm{sub}}(\delta u)$ emerging as an effective potential for the matrix model perturbed near $u^*$. As discussed above, the Cardy-like limit reproduces the dimensional reduction of the theory as the thermal $S^1$ is sent to zero. For generic non-zero $\tilde\Delta$ the whole KK tower of modes for the matter fields becomes massive and gets lifted as $r\to 0$, while a zero-mode survives in the vector multiplet and a gapped $3d$ pure CS gauge theory emerges. When the size of the $S^3$ is much larger than the size of the $S^1$, correlators at two different points on $S^3$ are exponentially suppressed and in the limit $r\to0$ the leading order description of the theory is captured only by CS contact terms \cite{Cassani:2021fyv}. In the language of localization the contribution from such contact terms are encoded in $I_{\mathrm{lead}}$ in \eqref{eq:cardyexp}. The sub-leading contribution $Z_{\mathrm{sub}}$ encodes the three-sphere partition function for the topological $\SU{N}_{\pm N}$ theory emerging in the reduction.

The saddle points configurations satisfy the saddle point equations
\begin{equation}
    \frac{\partial S_{4d}}{\partial u_i} \overset{!}{=} 0 \qquad i = 1,\dots, N-1.
\end{equation}
While in principle one needs to solve the saddle point equations derived from (\ref{eq:s_eff}), it is more convenient to compute the leading order saddle-point equations from the leading-order effective action, as for $r \to 0$ the saddles will converge to the leading order ones as discussed in \cite{GonzalezLezcano:2020yeb}. 

Employing modular properties of the elliptic gamma functions, Jacobi functions and $q$-Pochhammer symbols presented in appendix \ref{sec:ell_fun}, we can derive the complete expansion of the effective action in $r$ from (\ref{eq:s_eff}) up to exponentially suppressed terms. The leading order term $\sim\mathcal{O}(r^{-2})$ receives contributions only from matter fields
\begin{equation}
   S_{\mathrm{lead}} = - \frac{\pi \mi}{3 \tau \sigma} \left((N-1)\sum_{a = 1}^{3}B_3\left(\{\Delta_a\}\right) \,+\, \sum_{a = 1}^{3}\sum_{i \neq j}^{N} B_3 \left(\{u_{ij} + \Delta_a\}\right)\right),
\end{equation} 
where we define $\{x\} = \{\tilde x\} + r\bar x\equiv \tilde x - \lfloor \tilde x \rfloor + r \bar x$ for any $x$ with $\tilde x \neq 0$.
A set of solutions to the saddle point equations
\begin{eqnarray}
    \label{eq:saddle_point}
    \frac{\partial S_{4d}}{\partial u_i} &&=\sum_{a = 1}^{3}\sum_{j = 1}^N \Big(B_2(\{u_{ij} + \Delta_a\}) \,-\, B_2(\{-u_{ij} + \Delta_a\}) \nonumber \\[5pt]
    &&+ B_2(\{-u_{Nj} + \Delta_a\}) \,-\, B_2(\{u_{Nj} + \Delta_a\})\Big) = 0
\end{eqnarray}
is the so-called family of $C$-center solutions \cite{ArabiArdehali:2019orz,Cabo-Bizet:2019osg}, organized accordingly to the presence of a discrete one-form symmetry, namely the center symmetry $\mathbb{Z}_N$, and its subgroups \cite{ArabiArdehali:2019orz,Cassani:2021fyv}
\begin{equation}
    u_j = \frac{m}{N} + \frac{\left\lfloor\frac{j - 1}{N/C}\right\rfloor - \frac{C - 1}{2}}{C} \qquad j = 1,\dots, N,
    \label{eq:holoC}
\end{equation}
where $m = 0,\dots  \frac{N}{C} - 1$ and $C$ a divisor of $N$.

In the following we will be interested only in the $C = 1$ case, corresponding to the saddle point reproducing the black hole entropy. For $C = 1$, we have $N$ holonomy configurations with all the holonomies packed at $u_i = \frac{m}{N}$ for fixed $m = 0,\dots,N-1$, contributing the same to the index in (\ref{eq:ind_gauge}). The $\log N$ degeneracy arising from these saddles in the entropy function ($\sim \log\mathcal{I}$) is due to the index being insensitive to the presence of global properties of the gauge group, namely it is unable to detect the action of the $\mathbb{Z}_N$ center symmetry, mapping different saddles into each other. 
Expanding the effective action near vanishing holonomies with ansatz
\begin{equation}
    \label{eq:BH_ansatz}
    u_i = r\lambda_i, \quad \quad u_i \in \left[-\frac{1}{2},\frac{1}{2}\right)
\end{equation}
the gauge terms and Pochhammer symbols combines to produce the measure for a three-sphere partition function in the reduction
\begin{eqnarray}
    \mathcal{I} && =\frac{(p;p)_\infty^{N-1} (q;q)_\infty ^{N-1}}{N!}
    \int \prod_{i=1}^{N-1} \differential u_i \prod_{i < j} \theta_0( u_{ij}) \theta_0(-u_{ij})\dots \underset{r \to 0}{\sim}
    \nonumber \\
    && \sim  \frac{\ee ^ { - \frac{ \pi \mi (\omega_1 + \omega_2) (N^2 - 1)}{12 r \omega_1 \omega_2}}} {N!} 
    \int \prod_{i=1}^{N-1} \frac{\differential \lambda_i}{\sqrt{- \omega_1\omega_2}} \frac{1}{\prod_{i < j}  \Gamma_h( \lambda_{ij})\Gamma_h( - \lambda_{ij})}\dots 
\end{eqnarray}
where the dots stand for the matter content, whose contribution depends on the details of the reduction.
Parameterizing the complex chemical potentials $\Delta_a$ as in equation \eqref{eq:chemicalparam} the matter terms contribute as 
\begin{equation}
    \exp\left(2\pi \mi \left((N-1)\sum_{a = 1}^{3}Q\left(\{\Delta_a\}\right) \,+\, \sum_{a = 1}^{3}\sum_{i \neq j}^{N} Q \left(\{r \lambda_{ij} + \Delta_a\}\right)\right) + \mathcal{O}\left(\ee^{-\frac{1}{r}}\right)\right)
\end{equation}
which can be expanded for small $r$, generating a leading term $\sim r^{-2}$ related to central charges $a,c$ of the theory and a quadratic Chern-Simons term in the holonomies of order $\mathcal{O}(1)$ in $r$
\begin{equation}
    \exp(2\pi \mi \left((N^2 - 1)\sum_{a = 1}^{3}Q\left(\{\Delta_a\}\right) \,+\, r^2\left(\sum_{a = 1}^{3} Q'' \left(\{\Delta_a\}\right)\right)\sum_{i < j}^{N}(\lambda_i - \lambda_j)^2  \right) + \dots),
\end{equation}
where $Q''(x)$ denotes the second derivative of $Q(x)$ with respects to its argument and the dots stands for negligible terms in the Cardy-like limit at most of linear order in $r$. By virtue of the reality of the adjoint representation, the generation of an FI term linear in the gauge holonomies is prevented.

Upon employing the constraint
\begin{equation}
    \sum_{a=1}^{3}\{\Delta_a\} = \tau + \sigma + \frac{3 + n_0}{2}, \quad n_0 = \pm 1,
\end{equation}
which follows from the constraint (\ref{eq:del_cons}), and the definition of $Q(\{\Delta_a\})$
\begin{eqnarray}
    Q(\{\Delta_a\}) =&& -\frac{B_3(\{\Delta_a\})}{6 \,\sigma\tau } \,+\, B_2(\{\Delta_a\}) \frac{(\sigma + \tau )}{4\, \sigma\tau } \,-\, B_1(\{\Delta_a\})\frac{ \left((\sigma + \tau )^2 + \sigma \tau \right)}{12\, \sigma  \tau } \nonumber \\ 
    && +\, \frac{\sigma }{24} \,+\, \frac{\tau }{24},
\end{eqnarray}
we get
\begin{eqnarray}
    \exp\Bigg( && - \frac{\pi \mi (N^2 - 1)}{r^2 \omega_1\omega_2}\prod_{a = 1}^{3}\left(\{\Delta_a\} - \frac{n_0 + 1}{2} \right) - \frac{\pi \mi n_0 N}{\omega_1\omega_2} \sum_{i=1}^{N}\lambda_i^2 \nonumber \\
    && +\,\frac{\pi \mi (\omega_1 + \omega_2)(N^2 - 1)}{12r \omega_1\omega_2}  - \frac{\pi \mi n_0 (N^2 - 1)(\omega_1^2 + \omega_2^2 + 3\omega_1\omega_2)}{12\omega_1\omega_2}  \dots \Bigg),
\end{eqnarray}
All in all, the index becomes
\begin{eqnarray}
    \label{eq:4d_ind_CS}
    \mathcal{I} =&& \exp \left( - \frac{\pi \mi (N^2 - 1)}{\sigma\tau}\prod_{a = 1}^{3}\left(\{\Delta_a\} - \frac{n_0 + 1}{2} \right) - \frac{\pi \mi n_0(N^2 - 1)(\omega_1^2 + \omega_2^2 + 3\omega_1\omega_2)}{12\omega_1\omega_2}\right)\cdot\nonumber \\
    &&\cdot \frac{1}{N!} 
    \int \prod_{i=1}^{N-1} \frac{\differential \lambda_i}{\sqrt{- \omega_1\omega_2}} \frac{\ee^{- \frac{\pi \mi n_0 N}{\omega_1\omega_2} \sum_{i=1}^{N}\lambda_i^2}}{\prod_{i < j}  \Gamma_h( \lambda_{ij})\Gamma_h( - \lambda_{ij})},
\end{eqnarray}
consistently with \cite{Ardehali:2021irq}, where the domain of integration of each $\lambda_i = \frac{u_i}{r}$ runs over $(-\infty,+\infty)$ as $S^1$ shrinks to $0$.

The partition function for a pure CS theory on a squashed three-sphere background can be evaluated exactly in terms of the constrained $\U{N}$ CS partition function
\begin{equation}
    \label{eq:CS_part_SU(N)}
    \frac{1}{N!} 
    \int \dd \Lambda \int \prod_{i=1}^{N} \frac{\differential \lambda_i}{\sqrt{- \omega_1\omega_2}} \frac{\ee^{- \frac{\pi \mi n_0 N}{\omega_1\omega_2} \sum_{i=1}^{N}\lambda_i^2 + 2\pi \mi \Lambda \sum_{j = 1}^{N}\lambda_j}}{\prod_{i < j}  \Gamma_h( \lambda_{ij})\Gamma_h( - \lambda_{ij})}
\end{equation}
giving
\begin{equation}
    \mathcal{Z}^{S^3}_{\SU{N}_{n_0 N}} = \mathcal{Z}^{S^3}_{\U{N}_{n_0 N}} \sqrt{- \mi n_0}= \ee^{\tfrac{\pi \mi n_0\left(N^2 - 1\right) \left(\omega_1^2 + \omega_2^2 + 3\omega_1\omega_2 \right)}{12\omega_1\omega_2}}.
\end{equation}
Therefore, taking into account the $N$ degeneracy of the $1$-center saddles due to the action of the $\mathbb{Z}_N$-center symmetry, the final contribution to the index is
\begin{equation}
\mathcal{I} = N \exp \left( - \frac{\pi \mi (N^2 - 1)}{\sigma\tau}\prod_{a = 1}^{3}\left(\{\Delta_a\} - \frac{n_0 + 1}{2} \right) + \mathcal{O}(r)\right).
\end{equation}
\subsection{Adding the Defect}
\label{sec:DefectCL}
The insertion of a maximal Gukov-Witten defect amounts, in the SCI, to couple the $4d$ theory with the $2d$ model (\ref{eq:index2dfinal}) as described in Section \ref{sec:davide}. The Cartan generators of the half-BPS algebra can be identified with the ones in $4d$ and by comparing \eqref{eq:NSNSindex} with \eqref{eq:ind}. The dictionary between the fugacities is the following \cite{Chen:2023lzq}
\begin{equation}
    \sigma=\tau_{2d},\quad\tau=\frac{\tau_{2d}}{2}-u_C,\quad \Delta_1=\frac{\tau_{2d}}{2}+2\chi-u_C,\quad \Delta_2=\frac{\tau_{2d}}{2}+z-2\chi,
\end{equation} 
where $\sigma,\tau,\Delta_1,\Delta_2$ are the usual fugacities for $\mathcal{N}=4$.
In a Cardy-like limit approach for the evaluation of the index we notice that the insertion of such a defect modifies the original effective action with an order $\mathcal{O}\left(\frac{1}{r}\right)$ term
\begin{equation}
    \label{eq:I_4d_2d}
    \mathcal{I} = \frac{1}{N!} \sum_{i = 1}^{N} \int \prod_{j = 1}^{N-1}\dd u_j \ee^{S_{4d}(u,\Delta,\tau,\sigma) + S_{\mathrm{2d},i}(u,\Delta,\tau,\sigma)},
\end{equation}
where
\begin{equation}
    \label{eq:s_2d}
    S_{2d,i}(u,\Delta,\tau,\sigma) = \sum_{j \neq i}^{N}\frac{\log\theta_0(-u_{ij} - \Delta_2 +\sigma;\sigma)}{\log\theta_0(-u_{ij} + \Delta_1 - \tau;\sigma)} + \frac{\log\theta_0(u_{ij}  + \Delta_3;\sigma)}{\log\theta_0(u_{ij};\sigma)},
\end{equation}
giving rise to subleading corrections to the contribution of the 4d theory and crucially leaving the saddle-point equations (\ref{eq:saddle_point}) unaffected. Therefore, we can still identify the combined black hole/probe D3-brane system in the gravitational theory with the holonomy saddle, associated to the sole black hole solution in the unperturbed 4d theory. This is perfectly consistent with the holographic dual picture, in which a probe regime for the backreaction of the D3-brane on the black hole background is considered. In this regime, the backreaction effects are negligible and the insertion of a probe D3-brane does not spoil the underlying black hole geometry.
More generally, this can be extended to regime of charges beyond the black hole one, where other gravitational saddles are expected to provide a dominant contribution to the gravitational path integral. When the unperturbed dual 4d theory is considered, their behavior is described by the other $C-$center solutions to the saddle points equations or equivalently, in a Bethe-Ansatz approach for the evaluation of the index, by the contributions arising from the Hong-Liu solutions to the Bethe-Ansatz equations \cite{Hosseini:2016cyf,Hong:2018viz}. The insertion of a probe D3-brane to such gravitational solutions does not spoil the background geometry and the identification between $C$-center saddle points, in the dual theory, and the combined gravitational system still holds. In section \ref{sec:BAE} we will discuss how the probe regime manifests in the Bethe-Ansatz approach. 

Let us evaluate the effective action (\ref{eq:s_2d}) in the Cardy-like limit near the vanishing holonomies configuration defined in (\ref{eq:BH_ansatz}). By employing the asymptotic expansion for $\theta_0(u;\sigma)$, listed in Appendix \ref{sec:ell_fun}, we get
\begin{eqnarray}
    S_{2d,i} \sim && +\frac{\pi \mi (N - 1) \left(-2( \tau + \sigma )\{\Delta _1\} + \sigma \sum_{a=1}^{3}\{\Delta _a\} + \tau\right)}{\sigma} \nonumber \\ 
    && + \frac{\pi \mi  (N - 1) \left(2\{\Delta_1\}(\{\Delta _1\} - 1) - \sum_{a = 1}^3\{\Delta _a\}(\{\Delta_a\} - 1)\right)}{\sigma} \nonumber \\
    && + \frac{\pi \mi(N - 1)  \tau (\tau + \sigma)}{\sigma}  - \frac{2 \pi \mi (N - 1)\left(\sum_{a = 1}^{3}\{\Delta_a\} - 1 - \tau - \sigma\right)}{\sigma}\sum_{j \neq i}^{N}\frac{u_{ij}}{N-1} \nonumber \\
    && - \sum_{j \neq i} \Bigg(\log{\left(1-e^{-\frac{2\pi\mi}{\sigma}u_{ij}}\right)\left(1-e^{-\frac{2\pi\mi}{\sigma}(1 - u_{ij})}\right)} \nonumber \\
    && + \log{\left(1 - \ee^{-\frac{2\pi\mi}{\sigma}\left(\sigma - u_{ij} + 1 - \{\Delta_2\}\right)}\right)\left(1-\ee^{-\frac{2\pi\mi}{\sigma}\left(\sigma - u_{ij} + \{\Delta_2\}\right)}\right)} \nonumber \\
    && - \log{\left(1-\ee^{-\frac{2\pi\mi}{\sigma}\left(\{\Delta_1\} -u_{ij} - \tau\right)}\right)\left(1-\ee^{-\frac{2\pi\mi}{\sigma}\left(1 - \{\Delta_1\} -u_{ij} - \tau\right)}\right)} \nonumber \\
    && + \log{\left(1-\ee^{-\frac{2\pi\mi}{\sigma}\left(u_{ij} \{\Delta_3\}\right)}\right)\left(1-\ee^{-\frac{2\pi\mi}{\sigma}(1 - u_{ij} - \{\Delta_3\})}\right)}\Bigg).
\end{eqnarray}
For general values of $\Delta_a = \tilde \Delta_a + r \bar\Delta_a$ all the logarithmic terms but one are suppressed in the Cardy-like limit, assuming $\tilde\Delta_a \neq 0$. 
Upon constraining the chemical potentials according to eq. (\ref{eq:del_cons}) and employing the $\SU{N}$ constraint \eqref{eq:suck} we can rewrite the effective action as
\begin{eqnarray}
    \label{eq:s_2d_cardy}
        S_{2d,i} = &&  \frac{2\pi \mi (N - 1)}{\sigma}\prod_{a = 2}^{3}\left( \{\Delta_a\} - n \right)+ \pi\mi n (N -1)  - 2\pi\mi n N \frac{\lambda_{i}}{\omega_2} \nonumber \\
        && - \sum_{j \neq i} \log{\left(1-\ee^{-2\pi\mi\frac{\lambda_{ij}}{\omega_2}}\right)},
\end{eqnarray}
where we defined $n = \frac{1 + n_0}{2}$.

Before moving on, we would like to compare our result with the one in the recent paper \cite{Cabo-Bizet:2023ejm}. In that paper the authors derived the Cardy-like expansion of the $2d$ system by introducing a regulator in the asymptotic expansion of $\theta_0(u_{ij};\sigma)$ in terms of polylogarithms
\begin{equation}
    \log \theta_0(u_{ij};\sigma) = \frac{1}{2\pi \mi \sigma}\sum_{r = 0}^{\infty} (-1)^r\frac{(2\pi \mi \sigma)^r}{r!} \left(B_r(1 - z_{ij})\mathrm{Li}_{2-r}(\ee^{2\pi\mi\varepsilon}) + B_r(z_{ij})\mathrm{Li}_{2-r}(\ee^{-2\pi \mi \varepsilon})\right),
\end{equation}
so to have a well-defined expression during the manipulations. As far as the leading behavior of the $2d$ system is concerned, this is a perfectly fine choice, as the regulator does not spoil the leading $\mathcal{O}\left(\frac{1}{r}\right)$ terms in the effective action and indeed we find perfect agreement between their and our result, up to this order. The effect of the regulator only shows up at finite order in $r$, altering subleading corrections arising from the $2d$ model and thus, possibly preventing a clear understanding of corrections to the large $N$ limit and an EFT interpretation of the Cardy-like limit approach.
The net effect of the regulator is to suppress the logarithmic finite contribution arising from the asymptotic expansion of $\log \theta_0 (u_{ij};\sigma)$. To properly retrieve subleading effects in this approach, one would need to properly implement the regulator also in the $4d$ theory and include extra $\mathcal{O}(\sigma^0)$ effects arising from the vector multiplet of the $4d$ matrix model, before sending it to zero.

As long as a probe regime for the backreaction of the D3-brane is considered in the dual theory, we expect the contribution arising from the reduction of the $2d$ defect not being able to alter the effective $3d$ theory arising from the reduction of the sole $4d$ $\mathcal{N} = 4$ SYM. In the language of localization this translates into obtaining a matrix model for the gauge holonomies, expanded near the saddle point, associated to a pure CS partition function as in (\ref{eq:4d_ind_CS}). 
As a consequence, subleading corrections to the index in the probe limit from the $2d$ system can arise only from the expansion $\Delta = \tilde\Delta + r\bar\Delta$ in
\begin{equation}
\frac{2\pi\mi (N - 1)}{\sigma}\prod_{a = 2}^{3}(\{\Delta_a\} - n),
\end{equation}
as we will explicitly show below.
Plugging (\ref{eq:s_2d_cardy}) into (\ref{eq:I_4d_2d}) we get
\begin{equation}
\begin{split}
    \mathcal{I} = & N\exp\Bigg(-\frac{\pi \mi (N^2 - 1)}{\sigma\tau} \prod_{a = 1}^{3}\left(\{\Delta_a\} - n \right) +  \frac{2\pi\mi (N - 1)}{\sigma}\prod_{a = 2}^{3}(\{\Delta_a\} - n)\\
    & \phantom{N \exp()}- \frac{\pi \mi n_0(N^2 - 1)(\omega_1^2 + \omega_2^2 + 3\omega_1\omega_2)}{12\omega_1\omega_2}\Bigg)\cdot\\
    &\cdot \frac{1}{N!} \int \dd \Lambda \int \prod_{i=1}^{N} \frac{\differential \lambda_i}{\sqrt{- \omega_1\omega_2}} \frac{\ee^{- \frac{\pi \mi n_0 N}{\omega_1\omega_2} \sum_{i=1}^{N}\lambda_i^2 + 2\pi \mi \Lambda \sum_{j = 1}^{N}\lambda_j}}{\prod_{i < j}  \Gamma_h( \lambda_{ij})\Gamma_h( - \lambda_{ij})} \sum_{i = 1}^{N}\frac{\ee^{-2\pi\mi n N \frac{\lambda_{i}}{\omega_2} + \pi\mi n (N -1)} }{\prod_{j\neq i}^{N}\left(1 - \ee^{-2\pi \mi \frac{\lambda_{ij}}{\omega_2}}\right)}.
\end{split}
\end{equation}
We see that the defect deforms the original CS partition function with an extra term. However, this deformation is only apparent. In fact, let us consider 
\begin{equation}
    \label{eq:vande}
    \sum_{i = 1}^{N}\frac{\ee^{-2\pi\mi n N \frac{\lambda_{i}}{\omega_2} + \pi\mi n (N -1)} }{\prod_{j\neq i}^{N}\left(1 - \ee^{-2\pi \mi \frac{\lambda_{ij}}{\omega_2}}\right)}
\end{equation}
and define $z_j = \ee^{-2\pi \mi \frac{\lambda_j}{\omega_2}}$. Focusing first on the case $n = 0$, eq. (\ref{eq:vande}) can be rewritten as
\begin{equation}
    \sum_{i = 1}^{N}\prod\limits_{j\neq i}^{N} \frac{z_j}{\left(z_j - z_i\right)} = \left(\sum\limits_{i = 1}^{N} (-1)^{N - i}\prod\limits_{j\neq i}^{N} z_j\prod\limits_{\substack{1\leq k < l \leq N\\
    k,l\neq i}} (z_k - z_l)\right)\left(\prod\limits_{1\leq i<j \leq N}(z_i - z_j)\right)^{-1}.
\end{equation}
The last product is simply the Vandermonde determinant up to a sign due to the reordering of all the columns, while the first parenthesis can be rewritten in terms of a sum of monomials of degree $N(N-1)/2$ of the form 
\begin{equation}
    \label{eq:num_vande}
    \sum\limits_{i = 1}^{N} (-1)^{N - i}\prod\limits_{j\neq i}^{N} z_j\prod\limits_{\substack{1\leq k < l \leq N\\
    k,l\neq i}} (z_k - z_l) = \sum_{i = 1}^{N}(-1)^{N - i}\sum_{\sigma \in S_{N-1}}\mathrm{sign}(\sigma)\prod_{k = 1}^{N-1}z_{\sigma(\mathfrak{I}^i_k) }^{N-i},
\end{equation}
where $\mathfrak{I}^i = \{1,2,\dots,i-1,i+1,\dots,N\}$ and $\mathfrak{I}^i_k$ is the $k$-th element of such set. 
Written in this way, expression (\ref{eq:num_vande}) is simply the Laplace expansion of the Vandermonde determinant with an extra sign due to the very same reordering of columns already mentioned in the denominator.
 Similarly, for $n = 1$ eq. (\ref{eq:vande}) can be rewritten as, constraining $\prod_{j = 1}^{N}z_j = 1$,
 \begin{equation}
    \label{eq:2d_eft_def}
    \ee^{\pi \mi (N - 1)}\sum_{i = 1}^{N}\frac{z_i^{N - 1}}{\prod_{j\neq i}^{N}\left(z_j - z_i\right)} = 1.
\end{equation}
where we used the $\SU{N}$ constraint $\prod_{j=1}^{N}z_j = 1$. Then, the final expression for the index of the $4d$-$2d$ combined system in the Cardy-like limit is
\begin{equation}
    \mathcal{I} = N\exp\left(-\frac{\pi \mi (N^2 - 1)}{\sigma\tau} \prod_{a = 1}^{3}\left(\{\Delta_a\} - n \right) +  \frac{2\pi\mi (N - 1)}{\sigma}\prod_{a = 2}^{3}(\{\Delta_a\} - n)\right).
    \label{eq:finalcardy}
\end{equation}
In section \ref{sec:BAE} we will see that this result perfectly agrees with the Bethe-Ansatz evaluation of the index in the case of collinear angular momenta $\tau = \sigma$.
\subsection{EFT Interpretation}
In \cite{Chen:2023lzq} an EFT interpretation, along the lines of \cite{Cassani:2021fyv, ArabiArdehali:2021nsx}, was given for the case of equal charges at leading order in $N$. Such an interpretation uses the $\U{1}$ gauge theory formulation of the defect to reconstruct its contribution to the entropy function in terms of $2d$ anomalies. At sub-leading order, in the absence of the defect, a purely topological gapped CS gauge theory emerges for the massless modes in the Kaluza-Klein reduction \cite{Cassani:2021fyv, ArabiArdehali:2021nsx}. This contirbution arises in the Cardy-like limit as a $4d$ supersymmetric partition function on the squashed $S^3$ \cite{GonzalezLezcano:2020yeb,Amariti:2020jyx,Amariti:2021ubd}. In order to complete this EFT interpretation in presence of the surface defect, it is necessary to include its effect in the $3d$ topological theory. Here we have proved that the CS partition function (\ref{eq:CS_part_SU(N)}) is left unchanged by the addition of the surface defect. In the following we will give an EFT interpratiotion of this result. 

The reduction of a GW defect wrapping the temporal $S^1$ produces a line defect in the effective $3d$ pure CS theory. In absence of the denominator $\prod_{j \neq i}^{N}(z_j-z_i)$ in (\ref{eq:2d_eft_def}), we see that, when $n = 1$, the index receives the following sub-leading contribution
\begin{equation}
    \frac{\ee^{\pi\mi (N -1)}}{N!}  \int \dd \Lambda \int \prod_{i=1}^{N} \frac{\differential \lambda_i}{\sqrt{- \omega_1\omega_2}} \frac{\ee^{- \frac{\pi \mi n_0 N}{\omega_1\omega_2} \sum_{i=1}^{N}\lambda_i^2 + 2\pi \mi \Lambda \sum_{j = 1}^{N}\lambda_j}}{\prod_{i < j}  \Gamma_h( \lambda_{ij})\Gamma_h( - \lambda_{ij})} \left(\sum_{i = 1}^{N} \ee^{-2\pi\mi N \frac{\lambda_{i}}{\omega_2}}\right)
\end{equation}
which defines an $N$-wounded anti-fundamental Wilson loop insertion in the partition function of a pure CS theory on a squashed three-sphere with (analytically-continued) squashing parameters $\omega_1,\omega_2$. More precisely, for purely imaginary squashing parameters $\omega_1 = \mi b$ and $\omega_2 = \mi b^{-1}$, we have 
\begin{equation}
    \label{eq:wils}
    \left(\sum_{i = 1}^{N} \ee^{-2\pi N b^{-1} \lambda_{i}}\right),
\end{equation}
which defines the insertion of a the Wilson loop
\begin{equation}
    W_\gamma(\lambda) = \Tr_R \exp(\lambda \oint |\dot{x}|\dd s),
    \end{equation}
with lenght
\begin{equation}
    \oint \dd s = 2 \pi N b^{-1}.
\end{equation}
Defining the ellipsoid metric on $S^3$ as
\begin{equation}
    ds^2 = b^2(dx_0^2 + dx_1^2) + \tilde b^{-2}(dx_2^2 + dx_3^2),
\end{equation}
with
\begin{equation}
    x_0 = \cos\theta\cos\phi, \quad x_1 = \cos\theta\sin\phi, \quad x_2 = \sin\theta\cos\chi, \quad x_3 = \sin\theta\sin\chi,
\end{equation}
equation (\ref{eq:wils}) describes a $1/2$ BPS $N$-wounded Wilson loop wrapping the $1$-cycle at fixed $\chi$ on the $T^2$ in $S^3$, parametrized by $\chi,\phi$ coordinates at $\theta = \pi/2$, as discussed in \cite{Tanaka_2012}. The appearance of an exactly $N$-wounded Wilson loop from the reduction of a Gukov-Witten surface is crucial for a couple of reasons. Firstly, in a pure CS theory at level-$N$ only the expectation values of $p N$-wounded (anti-)fundamental Wilson loops, with $p \in \mathbb{Z}$, are non-vanishing. This can be seen more easily in the case of collinear angular momenta, for which $\tau = \sigma$ and $b=1$. In this case the insertion of a $pN$-wounded Wilson loop in the partition function of a pure CS theory gives
\begin{equation}
    \ee^{\mi \pi (N - 1)}Z^{\SU{N}_{\pm N}}_{W_{-N}} = N Z_{CS}^{\SU{N}_{\pm N}}.
\end{equation}
A derivation of this result is presented in Appendix \ref{sec:wils_loop}.
Secondly, we expect a symmetry between the $n = 0$ case, where the Wilson loop is not present, and the $n = 1$ case. In fact, the $4d$ index cannot detect global properties of the gauge groups and thus it is insensitive with respect to the action of the $\mathbb{Z}_N$ center symmetry. In addition, the insertion of a maximal Gukov-Witten defect introduces sub-leading corrections and thus it cannot alter this property, as discussed before for the saddle-point equations. Therefore, only an $N-$wounded Wilson loop is consistent with the case $n = 0$, being uncharged under the $\mathbb{Z}_N$ symmetry.
Let us now reintroduce and discuss the term $\prod_{j\neq i}(z_j - z_i)$.
As discussed before, the holographic counterpart of a maximal Gukov-Witten defect is described by a D3-brane in the probe limit. For this reason we expect the contribution of the defect not being able to alter the EFT emerging from the reduction of the theory along the thermal $S^1$. This manifests in the presence of
\begin{equation}
    -\log \theta_0(u_{ij};\sigma)
\end{equation}
in the $2d$ model describing the defect, which can be interpreted as a counter-term suppressing the effects of the Wilson loop emerging in the effective $3d$ pure CS theory. It would be interesting to study the fate of the counter-term for other GW defects in regimes where backreaction effects of the probe D3-brane are not necessarily negligible.


\section{Bethe Ansatz Approach}
\label{sec:BAE}
Motivated by the results just obtained, in this section we provide a derivation of the index in presence of the maximal GW defect using the BA approach. 
This technique was originally used in \cite{Benini:2018ywd}, following the derivation of \cite{Closset:2017bse}, in order to provide a derivation of the black-hole entropy at large $N$ beyond the Cardy-like regime. 
The result has been shown to be in perfect agreement with the one found by saddle-point approximation in the Cardy-like limit \cite{GonzalezLezcano:2020yeb,ArabiArdehali:2019orz}. 
We show that the agreement survives also in the presence of the maximal GW surface defect.


\subsection{The Bethe-Ansatz Formula}
\label{sec:BAAAF}
Here we review the BA formula \cite{Benini:2018mlo} in the context of $4d$ $\mathcal{N}=4$ SYM theory \cite{Benini:2018ywd}, for equal angular momenta
\begin{equation}
    \tau = \sigma \equiv \omega.
    \label{eq:equalcharges}
\end{equation}
We start by rewriting the SCI \eqref{eq:ind_gauge} in a more convenient way for the forthcoming discussion
\begin{equation}
    \label{symintind}
    \mathcal{I} = 
    \kappa_N \oint_{\mathcal{R}} \dd[N-1]{u}
    \mathcal{Z}_{4d}(u; \omega, \Delta)\,,
\end{equation}
where the integral is taken over the region
\begin{equation}
    \mathcal{R} = \left\{
        (u_1,\dots,u_{N-1}) \in \C^{N-1} \, | \, 
        \Re u_i \in \left[0,1\right]\,, \Im u_i =0\,, \forall \, i=1,\dots,N-1
    \right\}
\end{equation}
and the gauge holonomies are constrained by \eqref{eq:suck}. 
Then the prefactor $\kappa_N$ is given by
\begin{equation}
    \kappa_N = \frac{1}{N!}
    \left(
    \frac{
        (\ee^{2 \pi \mi \omega};\ee^{2 \pi \mi \omega})_{\infty}^2 \,
        \widetilde{\Gamma}(\Delta_1; \omega, \omega) \,
        \widetilde{\Gamma}(\Delta_2; \omega, \omega)
    }{
        \widetilde{\Gamma}(\Delta_1+\Delta_2; \omega, \omega) \,
    }
    \right)^{N-1},
\end{equation}
with the usual definitions of the $q$-Pochhammer symbol \eqref{poch} and of the elliptic gamma function \eqref{gamma}.
The integrand in \eqref{symintind}
\begin{equation}
    \mathcal{Z}_{4d}(u; \omega, \Delta) = 
    \prod_{i=1}^{N} \, \prod_{i \ne j=1}^{N}
    \frac{
        \widetilde{\Gamma}(u_{ij}+\Delta_1; \omega, \omega) \,
        \widetilde{\Gamma}(u_{ij}+\Delta_2; \omega, \omega) 
    }{
        \widetilde{\Gamma}(u_{ij}+\Delta_1+\Delta_2; \omega, \omega) \,
        \widetilde{\Gamma}(u_{ij}; \omega, \omega)
    }\,.
\end{equation}
It is important to stress that, in order to have a plethystic expansion of the elliptic functions, we need to restrict to a certain region of chemical potentials \cite{Benini:2018mlo}
\begin{equation}
    \label{dom}
    \mathcal{B} = \left\{
        \omega, \Delta \in \C \, | \,  
        0 < \Im \Delta < 2 \Im \omega
    \right\}\,,
\end{equation}
with $\Delta=\Delta_1,\,\Delta_2,\,\Delta_1+\Delta_2$. Then, once we have computed the index, we can eventually analytically continue the result outside $\mathcal{B}$.\\
The BA operators $Q_i$ are defined as
\begin{equation}
    \label{symbaos}
    Q_i(u; \omega, \Delta) \coloneqq 
    \ee^{2 \pi \mi (\lambda + 3 \sum_{j} u_{ij})}
    \prod_{j=1}^{N}
    \frac{
        \theta_0(u_{ji}+\Delta_1; \omega) \,
        \theta_0(u_{ji}+\Delta_2; \omega) \,
        \theta_0(u_{ji}-\Delta_1-\Delta_2; \omega)
    }{
        \theta_0(u_{ij}+\Delta_1; \omega) \,
        \theta_0(u_{ij}+\Delta_2; \omega) \,
        \theta_0(u_{ij}-\Delta_1-\Delta_2; \omega)
    }\,,
\end{equation}
where $i=1,\dots,N$ and the function $\theta_0$ is defined in \eqref{theta}. 
These operators, written for $\U{N}$ gauge symmetry, are restricted to the case of $\SU{N}$ by the action of the "Lagrange multiplier" $\lambda$.
A crucial property of $Q_i$ is that they shift the integrand in \eqref{symintind} as 
\begin{equation}
    \label{baoshift}
    \mathcal{Z}_{4d}(u-\delta_i \omega) = 
    \frac{Q_i}{Q_N} \, \mathcal{Z}_{4d}(u)\,,
    \qquad
    \forall \; i=1,\dots,N-1\,,
\end{equation}    
where 
\begin{equation}
    u-\delta_i\omega = 
    (u_1,\dots,u_i-\omega,\dots,u_{N-1},u_N+\omega)\,.
\end{equation}
Moreover, these operators are doubly periodic, i.e. they are invariant under the shifts
\begin{equation}
    u_i \mapsto u_i + m + n \omega \, ,
    \quad
    \forall \, m, n \in \Z \, ,
    \quad
    \forall \, i=1,\dots, N-1 \,.
\end{equation}
As we mentioned above, we can use \eqref{baoshift} to rewrite the integral representation \eqref{symintind} as
\begin{equation}
    \mathcal{I} = 
    \kappa_N \oint_{\mathcal{C}} \dd^{N-1}u \;
    \frac{
        \mathcal{Z}_{4d}(u; \omega, \Delta)
    }{
        \prod_{i=1}^{N} \left( 1 - Q_i(u; \omega, \Delta)\right)
    }\,,
\end{equation}
where now we are integrating over the contour $\mathcal{C}$ encircling the region
\begin{equation}
    \label{intreg}
    \mathcal{A} = \left\{
        u \in \C^{N-1} | 
        \Re u_i \in \left[0,1\right]\,,
        -\!\Im \omega < \Im u_i < 0, \forall \, i\!=\!1,\dots,N\!-\!1
    \right\}.
\end{equation}
At this point we can apply the residue theorem and recognize in the zeros of the denominator the only poles that really contribute.
In fact, a priori we should also consider those poles that come from $\mathcal{Z}_{4d}$. 
However, it turns out that, for each pole coming from the gamma functions inside $\mathcal{Z}_{4d}$, either there is a zero of the denominator of some $Q_i$ with higher multiplicity (thus canceling the pole), or such pole is outside $\mathcal{A}$ in \eqref{intreg} and thus cannot contribute.
Then the poles are obtained by the solutions of the set of trascendental equations
\begin{equation}
    \label{baes}
    Q_i(u;\omega,\Delta)=1 \,,
    \quad
    \forall \, i=1,\dots,N\,,
\end{equation}
the so called Bethe Ansatz Equations (BAEs).

We are almost ready to write the BA formula but first we need to clarify two aspects.
Firstly, due to the double periodicity of the operators $Q_i$, we can solve the BAEs on $N-1$ copies of the complex torus with modular parameter $\omega$. 
This means that the solutions can be grouped into a finite number of equivalence classes $[\hat{u}_i]$ such that
\begin{equation}
    \label{torus}
    \hat{u}_i \sim \hat{u}_i + 1 \sim \hat{u}_i + \omega\,,
    \quad
    \forall \; i=1,\dots,N\,,
    \qquad
    \sum_{i=1}^{N} \hat{u}_i = 0 \mod \left(\,\Z + \omega \Z\,\right).
\end{equation}
Secondly, as discussed in the Appendix C of \cite{Benini:2018mlo}, among all the solutions of the BAEs, there is the subset of all those solutions that are fixed by a non-trivial element of the Weyl group of $\SU{N}$.
It turns out that the integrand function $\mathcal{Z}_{4d}$ is such that the contributions from this subset sum up to zero and thus can be discarded.
These two clarifications bring us to define the set
\begin{equation}
    \label{mbae}
    \mathcal{M}_{\text{BAE}} \coloneqq 
    \left\{
        [\hat{u}] \in \mathcal{A} \; | \;
        Q_i([\hat{u}]; \omega, \Delta)=1 \, , 
        w \cdot [\hat{u}] \ne [\hat{u}]\, , 
        \; \forall \, i=1,\dots,N \, , \; \forall \, w \in S_N
    \right\}
\end{equation}
and finally the BA formula is given by
\begin{equation}
    \label{basym}
    \mathcal{I} = 
    \kappa_N
    \sum_{\hat{u} \in \mathcal{M}_{\text{BAE}}} 
    \mathcal{Z}_{4d}(\hat{u}; \omega, \Delta) \,
    H(\hat{u}; \omega, \Delta)^{-1}\,,
\end{equation}
where $H^{-1}$ is the inverse of a Jacobian due to the change of variables in the integral
\begin{equation}
    \label{jac}
    H(\hat{u}; \omega, \Delta) = \det \left(\,
        \frac{1}{2 \pi \mi}
        \left.
        \frac{
            \partial(Q_1,\dots,Q_N)
        }{
            \partial(u_1,\dots,u_{N-1},\lambda)
        }
        \right|_{\hat{u}}\,
    \right)\,.
\end{equation}
Unfortunately the full set of solutions of the BAEs \eqref{baes} has not been found yet.
However, a subset of solutions is known \cite{Benini:2018ywd,Hong:2018viz,Hosseini:2016cyf}.
Within this subset, one solution, also known as basic solution,
\begin{equation}
    \label{basic}
    \hat{u}_i = \bar{u}-\frac{\omega}{N} \, i \;, 
    \quad \text{with $\bar{u}$ such that} \quad
    \sum_{i=1}^{N} \hat{u}_i=0 \mod \left(\Z + \omega \Z\right),
\end{equation}
reproduces the leading contribution to the index whose logarithm matches with the entropy function of the dual $5d$ rotating black hole solution in holography.
This result was first obtained in \cite{Benini:2018ywd} and then improved in \cite{GonzalezLezcano:2020yeb}
\begin{equation}
    \label{symbasicind}
    \begin{aligned}
         \mathcal{I}\big|_{\text{basic}} 
        &= \left. N N! \, \kappa_N \, \mathcal{Z}_{4d} \, H^{-1} \right|_{\text{basic}} = \\
        &= \exp \left(
            -\frac{\pi \mi}{\omega^2} \, N^2 \,
            \prod_{a=1}^{3} \left( \, \left\{\Delta_a\right\}_{\omega}-n \, \right)
            + \log N 
            + O(N^0)
            \right)\,,
    \end{aligned}
\end{equation}
where $n=\tfrac{1+n_0}{2}$, $n_0=\pm1$, the function $\{\,\cdot\,\}_\omega$ is defined as
\begin{equation}
    \label{curly}
    \left\{\Delta\right\}_\omega \coloneqq 
    \Delta + m 
    \quad \text{ such that } \quad 
    m \in \Z 
    \; \text{ and } \;
    0 > \Im \, \frac{\Delta + m}{\omega} > \Im \frac{1}{\omega}
\end{equation}
and the auxiliary chemical potential $\Delta_3$ in \eqref{symbasicind} such that
\begin{equation}
    \label{constr}
    \left\{\Delta_1\right\}_\omega +
    \left\{\Delta_2\right\}_\omega + 
    \left\{\Delta_3\right\}_\omega =    
    2 \omega + \frac{3+n_0}{2}\,.
\end{equation}
We added an extra pre-factor $N\cdot N!$ representing the multiplicity of the basic solution, that can be justified as follows. 
As we mentioned, we consider only those solutions that are not fixed by any non-trivial element of $S_N$, but this implies that there is a multiplicity factor $N!$ related to the Weyl group action on each solution. 
Moreover we observe that, since the BAEs and the index depend only on the differences $u_{ij}$, we can shift $\bar{u} \mapsto \bar{u}+i/N$ into \eqref{basic}, with $i=0,\dots,N-1$, to obtain a set of $N$ inequivalent solutions giving the same contribution to the index. 
The shift is chosen in such a way that the constraint \eqref{torus} always holds. 
However, this implies that there is another multiplicity factor $N$ related to these shifts. 
This is the reason why the total multiplicity for the basic solution is $N\cdot N!$.\\
Finally we can analytically continue \eqref{symbasicind} outside the region \eqref{dom}, so to extend the result to any $\Delta \!\in\! \C$ such that 
\begin{equation}
    \label{stokes}
    \Im \, \frac{\Delta}{\omega} \notin \Z \times \Im \, \frac{1}{\omega} \,,
\end{equation}
because $\{\,\cdot\,\}_\omega$ is not defined on these lines, denoted as Stokes lines in \cite{Benini:2018ywd}.

Before continuing to the next sub-section, we make a further comment on the matching between the entropy function obtained from the Cardy-like limit of the $4d$ SCI and the BA approach. Such matching extends beyond the functional agreement and it relates the saddle holonomies of \eqref{eq:holoC} with the Hong-Liu solutions \cite{Hong:2018viz}. A complete discussion of such matching has been done in \cite{ArabiArdehali:2019orz}. We have seen from the saddle-point analysis in sub-section \ref{sec:DefectCL}, that the holonomy saddles giving rise to the $5d$ black hole are not modified in presence of the defect. Here we wonder if a counterpart of this behavior is realized in the BA approach. In the next sub-section we will give an affirmative answer to this question by studying the pole of the BA formula in presence of the defect.


\subsection{The Bethe-Ansatz Formula in Presence of the Defect}
\label{ssbadef}
The addition of the GW defect modifies the SCI as follows
\begin{equation}
    \label{defind}
    \mathcal{I} = 
    \int_{\mathcal{R}} 
    \dd^{N-1}u \;
    \mathcal{Z}_{4d}(u; \omega, \Delta) \, 
    \mathcal{Z}_{2d}(u; \omega, \Delta)\,,
\end{equation}
where the $2d$ contribution is given by
\begin{equation}
    \label{2dint}
    \mathcal{Z}_{2d} = 
    \sum_{i=1}^{N} \exp \left(
    \sum_{ i \ne j = 1}^{N} 
    \left(
    \log \frac
    {\theta_0(-u_{ij}\!-\!\Delta_2\!+\!\omega; \omega)}
    {\theta_0(-u_{ij}\!+\!\Delta_1\!-\!\omega; \omega)} + 
    \log \frac
    {\theta_0(u_{ij}\!-\!(\Delta_1\!+\!\Delta_2)\!+\!2\omega; \omega)}
    {\theta_0(u_{ij}; \omega)}  
    \right)
    \right)\,.
\end{equation}
In light of the discussion at the end of sub-section \ref{sec:BAAAF}, we study the feasibility of the BA approach in this new situation by inserting $\mathcal{Z}_{2d}$ in \eqref{basym} as
\begin{equation}
    \label{badef}
    \mathcal{I} \stackrel{?}{=} 
    \kappa_N
    \sum_{\hat{u} \in \mathcal{M}_{\text{BAE}}} 
    \mathcal{Z}_{4d}(\hat{u}; \omega, \Delta) \,
    \mathcal{Z}_{2d}(\hat{u}; \omega, \Delta) \,
    H(\hat{u}; \omega, \Delta)^{-1}\,.
\end{equation}
The answer will turn out to be affirmative even if one has to be cautious.
The proof of the BA formula guarantees that there is no contributing pole from $\mathcal{Z}_{4d}$ but a priori we cannot be sure that this still holds when we have $\mathcal{Z}_{4d}\,\mathcal{Z}_{2d}$.
In fact, the presence of poles in $\mathcal{Z}_{2d}$ within the region $\mathcal{A}$ previously defined in \eqref{intreg}, could potentially spoil the result. We rewrite the region in \eqref{intreg} as
\begin{equation}
    \label{intreg1}
    \mathcal{A} = \left\{
        u \in \C^{N-1} | 
        \Re u_i \in \left[0,1\right]\,,
        -\!\Im \omega < \Im u_{ij} < \Im \omega, \forall \, i,j\!=\!1,\dots,N\!-\!1
    \right\},
\end{equation}
in order to make the next discussion more intuitive.

Recall that, according to \cite{Kharchev_2015}, the zeros of the $\theta_0$ functions are given by
\begin{equation}
    \label{thzero}
    \theta_0(u; \omega)=0 \quad \iff \quad u=m+n\omega\,, \quad \forall \, m,n \in \Z \,.
\end{equation}
Therefore, from the definition \eqref{badef}, there is a pole whenever
\begin{equation}
    \begin{aligned}
        & (\mathrm{A}) \quad \theta_0(u_{ij};\omega) = 0 \\
        & (\mathrm{B}) \quad \theta_0(-u_{ij}+\Delta_1;\omega) = 0
    \end{aligned}
    \qquad \iff \qquad
    \begin{aligned}
        &u_{ij} = m \omega \, ,\\
        &u_{ij} = \Delta_1 + m \omega 
    \end{aligned}
    \qquad \forall \, m \in \Z\,.
\end{equation}
By rewriting 
\begin{equation}
    \prod_{i=1}^{N} \, \prod_{i \ne j=1}^{N} \frac{1}{\widetilde{\Gamma}(u_{ij};\omega,\omega)}=\prod_{i=1}^{N} \, \prod_{i \ne j=1}^{N} \theta_0(u_{ij};\omega)
\end{equation}
in $\mathcal{Z}_{4d}$, we see that each pole of type \!$(\mathrm{A})$ is cancelled by the corresponding zero in $\mathcal{Z}_{4d}$. For poles of type \!$(\mathrm{B})$, we start by noticing that, since $\Delta_1 \in \mathcal{B}$ \eqref{dom}, only those poles with $m=0,-1,-2$ can lie inside the region $\mathcal{A}$.
Secondly, we recall that inside $Q_j$ there is a term of the form
\begin{equation}
    \frac{1}{\theta_0(u_{ji}+\Delta_1;\omega)}\,,
    \label{eq:mhhhh}
\end{equation}
hence its denominator is zero whenever
\begin{equation}
    u_{ij} = \Delta_1 + n \omega\,, \quad \forall \, n \in \Z\,.
\end{equation}
However, some of these zeros were formerly used to take care of the corresponding poles coming from the gamma function in $\mathcal{Z}_{4d}$.
In fact, 
\begin{equation}
    \widetilde{\Gamma}(u_{ji}+\Delta_1; \omega, \omega) = 
    \prod_{\ell=0}^{\infty} 
        \Bigg(\frac{
            1-
            \ee^{2 \pi \mi (\ell+2) \omega} 
            \ee^{2 \pi \mi (u_{ij}-\Delta_1)}
        }{
            1-
            \ee^{2 \pi \mi \ell \omega} 
            \ee^{-2 \pi \mi (u_{ij}-\Delta_1)}
        }  \Bigg)^{\ell+1}
\end{equation}
has a pole of multiplicity $\ell+1$ whenever
\begin{equation}
    u_{ij} = \Delta_1 + \ell \omega \,,
    \quad \forall \, \ell \in \N\,.
\end{equation}
Poles with $\ell\ne0$ are outside $\mathcal{A}$ because $\Delta_1 \in \mathcal{B}$ of \eqref{dom}. On the other hand, the pole $u_{ij}=\Delta_1$ lies inside $\mathcal{A}$, thus we need the corresponding zero of the denominator of \eqref{eq:mhhhh} with $n=0$ to cancel it.
Therefore, on one hand we can use the zeros of the denominator of $Q_j$ with $n=-1,-2$ to cancel the corresponding poles in $\mathcal{Z}_{2d}$ with $m=-1,-2$. On the other, we lack a further zero to cancel the pole with $m=0$.

This is not the end of the story yet. In the proof of the BA formula in absence of the defect, we chose to restrict the integral to the domain $\mathcal{B}$ in order to have a plethystic expansion of the elliptic functions, and only at the end of computation we extended the resulting index outside this domain by analytic continuation.
In the same fashion, here we can first restrict the integration on a smaller domain for $\Delta_1$, 
\begin{equation}
    \label{dom2}
    \mathcal{B}\,' = \left\{
        \Delta_1 \in \C \, | \,  
        \Im \omega < \Im \Delta_1 < 2 \Im \omega
    \right\} \,,
\end{equation}
such that the pole of $\mathcal{Z}_{2d}$ with $m=0$ pops out of $\mathcal{A}$. This allow us to apply the BA formula because $\mathcal{Z}_{2d}$ will not bring new poles contributing to the integral.
Finally, we will extend the result outside $\mathcal{B}\,'$ by analytic continuation, when possible.


\subsection{Contribution of the Basic Solution}

We are now ready to evaluate the index \eqref{2dint} on the basic solution of the BAEs \eqref{basic} in presence of the GW defect. Such computation aims to generalize the result of \cite{Chen:2023lzq} which is restricted to the case of equal chemical potentials 
\begin{equation}
    \Delta_1=\Delta_2=\Delta_3=\frac{2\omega-1}{3}.
\end{equation} 
Here we will consider the case of arbitrary $\omega$ and $\Delta_{1,2,3}$, with the constraint \eqref{constr}.
As a starting point, we compute the following sum
\begin{equation}
    \label{piece}
    \sum_{i\neq j=1}^{N} \log \theta_0(\pm \hat{u}_{ij} + v; \omega)\,
\end{equation}
Firstly, by using the modular transformations
\begin{equation}
    \label{modular}
    \theta_0(u;\omega) = \ee^{\pi \mi B(u;\omega)} \theta_0 \left(\frac{u}{\omega}; -\frac{1}{\omega}\right)\,,
    \quad
    B(u;\omega) = -\frac{u^2}{\omega}-\frac{u}{\omega}+u-\frac{\omega}{6}-\frac{1}{6\omega}+\frac{1}{2}
\end{equation}
we obtain
\begin{equation}
    \text{\eqref{piece}} = 
    \pi \mi \sum_{ i \ne j = 1}^{N} B \left(\pm \hat{u}_{ij}\!+\!v; \omega\right) +
    \sum_{ i \ne j = 1}^{N} 
    \log \theta_0 \left(\frac{\pm \hat{u}_{ij}\!+\!v}{\omega}; -\frac{1}{\omega}\right)
    \equiv
    \phi\left(v,\omega\right) + \varphi\left(v,\omega\right)\,.
    \label{eq:aiutoPietro}
\end{equation}
Secondly, focusing on $\varphi(v,\omega)$ of the sum above and recalling the definitions \eqref{poch}, \eqref{theta}, we get the following expression
\begin{equation}
    \varphi(v,\omega) = 
    \sum_{ i \ne j = 1}^{N} \sum_{m=0}^{\infty} 
    \log
    \left( 1 - \tilde{w} \, \tilde{h}^m \,
    \left(\frac{\tilde{z}_i}{\tilde{z}_j}\right)^{\pm 1} 
    \right) \!
    \left( 1 - \tilde{w}^{-1} \, h^{m+1} \,
    \left(\frac{\tilde{z}_j}{\tilde{z}_i}\right)^{\pm 1} 
    \right)\,, 
\end{equation}
where, for compactness, we have defined
\begin{equation}
    \label{fug}
    \tilde{z}_j \equiv \ee^{\frac{2\pi \mi }{\omega} \hat{u}_j}\,, \qquad        
    \tilde{w}_a \equiv \ee^{\frac{2\pi \mi }{\omega} v}\,,
    \qquad
    \tilde{h} \equiv \ee^{-\frac{2\pi \mi}{\omega}}\,.
\end{equation}
By the Taylor expansion of the logarithm, we obtain
\begin{equation}
    \varphi(v,\omega) = -
    \sum_{ i \ne j = 1}^{N} 
    \sum_{m=0}^{\infty} 
    \sum_{n=1}^{\infty} 
    \frac{1}{n} 
    \left( \,
    \tilde{w}^n \, \tilde{h}^{mn} \,
    \left(\frac{\tilde{z}_i}{\tilde{z}_j}\right)^{\pm n} + 
    \tilde{w}^{-n} \, \tilde{h}^{(m+1)n} \,
    \left(\frac{\tilde{z}_j}{\tilde{z}_i}\right)^{\pm n} \,
    \right)\,.
\end{equation}
Now we evaluate explicitly the following sums on the basic solution \eqref{basic} 
\begin{equation}
    A_n \equiv \sum_{ i \ne j = 1}^{N} \left(\frac{\tilde{z}_i}{\tilde{z}_j}\right)^n \;,
    \qquad
    B_n \equiv \sum_{ i \ne j = 1}^{N} \left(\frac{\tilde{z}_j}{\tilde{z}_i}\right)^n \;,
\end{equation}
which amount to
\begin{equation}
    A_n = B_n = 
    \begin{cases}
        -1 & n \ne 0 \quad \text{mod} \, N \\
        N-1 & n = 0 \quad \text{mod} \, N\,.
    \end{cases}
\end{equation}
By summing over $m$, we can reorganize $\varphi(v,\omega)$ in a simpler form
\begin{equation}
    \label{pdio}
    \varphi(v,\omega) = 
    \sum_{n=1}^{\infty} \frac{1}{n} \,
    \frac{\tilde{w}^n + \tilde{w}^{-n} \, \tilde{h}^n}{1-\tilde{h}^n} -
    \sum_{n=1}^{\infty} \frac{1}{n} \,
    \frac{\tilde{w}^{Nn} + \tilde{w}^{-Nn} \, \tilde{h}^{Nn}}{1-\tilde{h}^{Nn}}
\end{equation}
where the first term is convergent if we take
\begin{equation}
\label{domain}
    \lvert \tilde{h} \rvert < \lvert \tilde{w} \rvert < 1
    \quad \iff \quad 
    \Im -\frac{1}{\omega} > \Im \frac{v}{\omega} > 0\,,
\end{equation}
while the second term, within the region \eqref{domain}, vanishes for large $N$.

The first three terms of $\mathcal{Z}_{2d}$, depending on $\Delta_1, \Delta_2, \Delta_1+\Delta_2$ respectively, have the same form of \eqref{piece} and thus we use the result \eqref{pdio} to compute them.
More precisely, we first apply the modular transformations \eqref{modular} and then, by the quasi-periodicity, 
\begin{equation}
\label{quasiper}
    \theta_0\left( \frac{u}{\omega};-\frac{1}{\omega} \right) = 
    \ee^{\pi \mi S(u; \,\omega)} 
    \theta_0\left( \frac{u}{\omega} -\frac{1}{\omega};-\frac{1}{\omega} \right) \,,
    \quad
    S(u;\omega) = 1 + \frac{2u}{\omega}\,,
\end{equation}
we shift by $-1/\omega$ the arguments of the two $\theta_0$ in the numerator of $\mathcal{Z}_{2d}$, that depend on $\Delta_2, \Delta_1+\Delta_2$ respectively.
This further step guarantees the same domain of convergence for each of these three terms. 
Hence these terms of $\mathcal{Z}_{2d}$ become
\begin{eqnarray}
    &1^{\text{st}}\,&:\quad
    + \pi \mi \sum_{ i \ne j = 1}^{N}        
    \left(
    B(-\hat{u}_{ij}-\Delta_2+\omega; \omega) +
    S(-\hat{u}_{ij}-\Delta_2; \omega)
    \right)
    + \order{N^0}\,, \nonumber
    \\
    &2^{\text{nd}}\,&:\quad
    + \pi \mi \sum_{ i \ne j = 1}^{N}
    \left(
    B(\hat{u}_{ij}-(\Delta_1+\Delta_2)+2\omega; \omega) + 
    S(\hat{u}_{ij}-(\Delta_1+\Delta_2); \omega) 
    \right)
    + \order{N^0}\,, \nonumber
    \\
    &3^{\text{rd}}\,&:\quad 
    - \pi \mi \sum_{ i \ne j = 1}^{N}
    B(-\hat{u}_{ij}+\Delta_1-\omega; \omega) 
    + \order{N^0}\,,
\end{eqnarray}
provided that the chemical potentials are taken in the domain of convergence
\usepgfplotslibrary{fillbetween}
\begin{figure}
    \centering
    \begin{tikzpicture}
    \begin{axis}[
        axis x line=middle,    
        axis y line=middle,   
        xlabel={$\text{Re}\, \Delta$}, 
        ylabel={$\text{Im}\, \Delta$},
        y label style={anchor=south},
        x label style={anchor=west},
        ticks=none,
        xmin=-2.2,xmax=2.2,
        ymin=-1,ymax=2.2,
    ]
        \plot[name path=Y,dashed,samples=100,domain=-2.2:2.2] {2*x+6};
        \plot[name path=Z,dashed,samples=100,domain=-2.2:2.2] {2*x+4};
        \plot[name path=A,thick,samples=100,domain=-2.2:2.2] {2*x+2};
        \plot[name path=B,thick,samples=100,domain=-2.2:2.2] {2*x};
        \plot[name path=C,dashed,samples=100,domain=-2.2:2.2] {2*x-2};
        \plot[name path=D,dashed,samples=100,domain=-2.2:2.2] {2*x-4};
        \addplot fill between[ 
            of = A and B,
            soft clip={domain=--2:2},
            every even segment/.style = {verdepetrolio,opacity=.2}
        ];
        \node [black] at (axis cs: -1,0) {\textbullet};
        \node [above left] at (axis cs: -1,0) {$-1$};
        \node [black] at (axis cs: 0.25,0.5) {\textbullet};
        \node [right] at (axis cs: 0.25,0.5) {$\omega$};
    \end{axis}
    \end{tikzpicture}
    \caption{The colored region represents the portion of the $\Delta$-complex plane where $\Delta_1, \Delta_2, \Delta_1 + \Delta_2$ live in order to have a convergent plethystic expansion. The dashed lines are defined in \eqref{stokes}.}
    \label{fig:domain}
\end{figure}
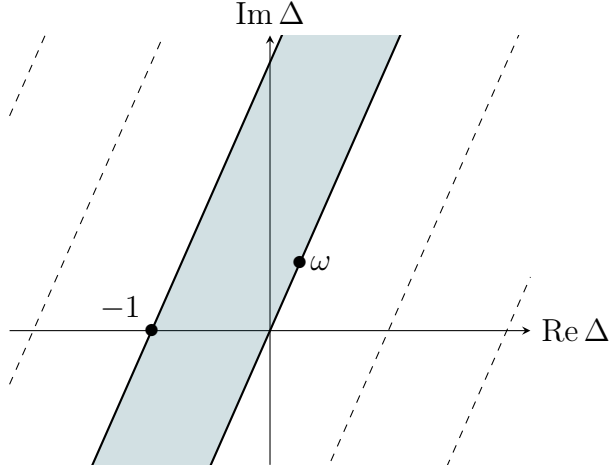
\begin{equation}
    \Delta_1, \, \Delta_2, \, \Delta_1 + \Delta_2 \in
    \mathcal{D} \equiv 
    \left\{ 
    \Delta \in \C \, \, : \, \,
    \Im - \frac{1}{\omega} > \Im \frac{\Delta}{\omega} > 0 
    \right\}\,,
\end{equation}
which is shown in Figure \ref{fig:domain}.
However, because of the quasi-periodicity of $\theta_0$, the $2d$ defect integrand is invariant under the shifts $\Delta_a \mapsto \Delta_a + n$, for any integer $n$.
This means that, if we introduce the function $[\,\cdot\,]_\omega$, defined as
\begin{equation}
    \label{squared}
    \left[\Delta\right]_\omega \coloneqq \Delta + n 
    \quad \text{such that} \quad 
    n \in \Z \quad \text{and} \quad \Im - \frac{1}{\omega} > \Im \, \frac{\Delta + n}{\omega} > 0\,,
\end{equation}
for any $\Delta \!\in\! \C$ such that
\begin{equation}
    \label{stokes2}
    \Im \, \frac{\Delta}{\omega} \notin \Z \times \Im \, \frac{1}{\omega}\,,
\end{equation}
we can extend the result to the whole complex plane by analytic continuation, with the exception of the lines \eqref{stokes2}, and get
\begin{eqnarray}
    \label{partial1-3}
    &1^{\text{st}}\,&:\quad
    + \pi \mi \sum_{ i \ne j = 1}^{N} 
    \left(
    B(-\hat{u}_{ij}-[\Delta_2]_{\omega}+\omega; \omega) + 
    S(-\hat{u}_{ij}-[\Delta_2]_{\omega}; \omega)
    \right)
    + \order{N^0}\,, \nonumber
    \\
    &2^{\text{nd}}\,&:\quad 
    + \pi \mi \sum_{ i \ne j = 1}^{N}
    \left(
    B(\hat{u}_{ij}-[\Delta_1+\Delta_2]_{\omega}+2\omega;\omega) + 
    S(\hat{u}_{ij}-[\Delta_1+\Delta_2]_{\omega}; \omega) 
    \right)
    + \order{N^0}\,, \nonumber
    \\
    &3^{\text{rd}}\,&:\quad 
    - \pi \mi \sum_{ i \ne j = 1}^{N}
    B(-\hat{u}_{ij}+[\Delta_1]_{\omega}-\omega; \omega) 
    + \order{N^0}\,,
\end{eqnarray}
We still have to compute the fourth term of $\mathcal{Z}_{2d}$ which requires a different approach
\begin{equation}
    \begin{aligned}
        4^{\text{th}}\,:\quad 
        &- 
        \sum_{ i \ne j = 1}^{N} 
        \sum_{m=0}^{\infty} 
        \log 
        \left(
        1 - \tilde{h}^m
        \left(\frac{\tilde{z}_i}{\tilde{z}_j}\right)
        \right)
        \left( 
        1 - \tilde{h}^{m+1}
        \left(\frac{\tilde{z}_j}{\tilde{z}_i}\right)
        \right) =
        \\
        = &- 
        \sum_{ i \ne j = 1}^{N} 
        \log 
        \left(
        1 - \left(\frac{\tilde{z}_i}{\tilde{z}_j}\right)
        \right) + 2 
        \sum_{ i \ne j = 1}^{N} 
        \sum_{m=1}^{\infty} 
        \log 
        \left( 
        1 -  \tilde{h}^m
        \left(\frac{\tilde{z}_i}{\tilde{z}_j}\right)
        \right)\,.
    \end{aligned}
    \label{fdmfdp}
\end{equation}
The second term in \eqref{fdmfdp}, in analogy with the previous computation, once evaluated on the basic solution, is given by
\begin{equation}
    2 
    \sum_{ i \ne j = 1}^{N} 
    \sum_{m=1}^{\infty} 
    \log 
    \left(
    1 -  \tilde{h}^m
    \left(\frac{\tilde{z}_i}{\tilde{z}_j}\right)
    \right)
    =
    \sum_{n=1}^{\infty} 
    \frac{1}{n} \, 
    \frac{\tilde{h}^n}{1-\tilde{h}^n} -
    \sum_{n=1}^{\infty} 
    \frac{1}{n} \, 
    \frac{\tilde{h}^{Nn}}{1-\tilde{h}^{Nn}} = \order{N^0}\,,
\end{equation}
at large $N$ and $\lvert \tilde{h} \rvert < 1$. For the first term in \eqref{fdmfdp}, notice that, since
\begin{equation}
    x^N - 1 = \prod_{k=1}^N \left( x - \ee^{2 \pi i \frac{k}{N}} \right)\,,
\end{equation}
by factorizing $x-1$, we have
\begin{equation}
    x^{N-1} + \dots + x + 1 = 
    \prod_{k=1}^{N-1} \left( x - \ee^{2 \pi i \frac{k}{N}} \right)\,,
\end{equation}
that, for $x=1$, becomes
\begin{equation}
    N = \prod_{k=1}^{N-1} \left( 1 - \ee^{2 \pi i \frac{k}{N}} \right)
\end{equation}
and finally we conclude that the contribution we are looking for amounts to
\begin{equation}
    \label{partial4}
    4^{\text{th}}\,:\quad - \log N + O(N^0)\,.
\end{equation}
Collecting our results \eqref{partial1-3} and \eqref{partial4}, we obtain
\begin{equation}
    \left.\mathcal{Z}_{2d}\right|_{\text{basic}} =
    \sum_{i=1}^N
    \exp
    \left(
    \Psi_i([\Delta_1]_\omega, [\Delta_2]_\omega, [\Delta_1+\Delta_2]_\omega)
    - \log N + \order{N^0}
    \right)\,,
\end{equation}
where we introduced the function
\begin{equation}
    \begin{aligned}
        &\Psi_i([\Delta_1]_\omega, [\Delta_2]_\omega, [\Delta_1+\Delta_2]_\omega) 
        = \\
        &= \pi \mi N \left(
        - 3 [\Delta_1]_\omega \!+\! [\Delta_2]_\omega \!+\! 3 [\Delta_1\!+\!\Delta_2]_\omega
        \right)
        - \frac{2 \pi \mi}{\omega} N \left( 
        [\Delta_1]_\omega \!+\! [\Delta_2]_\omega \!-\! [\Delta_1\!+\!\Delta_2]_\omega
        \right)\! \sum_{i \ne j = 1}^N \hat{u}_{ij} + \\
        &+ \frac{\pi \mi}{\omega} N \left(
        [\Delta_1]_\omega \!-\! [\Delta_2]_\omega\!-\! [\Delta_1\!+\!\Delta_2]_\omega
        \right) +
        \frac{\pi \mi}{\omega} N \left(
        [\Delta_1]_\omega^2 \!-\! [\Delta_2]_\omega^2 \!-\! [\Delta_1\!+\!\Delta_2]_\omega^2 
        \right)\,.
    \end{aligned}
\end{equation}
Here we can distinguish two cases:
\begin{equation}
    \left[\Delta_1 + \Delta_2\right]_\omega = 
    \begin{cases}
        \left[\Delta_1\right]_\omega + 
        \left[\Delta_2\right]_\omega \quad
        &\text{for} \quad
        \Im - \frac{1}{\omega} > 
        \Im \frac{
        \left[\Delta_1\right]_\omega +
        \left[\Delta_2\right]_\omega}{\omega} > 0 \\
        \left[\Delta_1\right]_\omega + 
        \left[\Delta_2\right]_\omega + 1 \quad
        &\text{for} \quad 
        \Im - \frac{2}{\omega} > 
        \Im \frac{
        \left[\Delta_1\right]_\omega +
        \left[\Delta_2\right]_\omega}{\omega} >
        \Im - \frac{1}{\omega}
    \end{cases}
\end{equation}
or equivalently, in a more compact form, 
\begin{equation}
    \left[\Delta_1 + \Delta_2\right]_\omega = 
    \left[\Delta_1\right]_\omega + \left[\Delta_2\right]_\omega +
    \frac{1-n_0}{2}
    \qquad \text{where} \qquad
    n_0 \coloneqq 
    \begin{cases}
        +1 \quad &\text{I} \\
        -1 \quad &\text{II} \;.
    \end{cases}
\end{equation}
We introduce the auxiliary chemical potential $\Delta_3$ constrained as in \eqref{eq:del_cons} with the further assumption \eqref{eq:equalcharges} and, because of the properties of $[\,\cdot\,]_\omega$,
\begin{equation}
    [\Delta+n]_\omega = [\Delta]_\omega \,,
    \qquad
    [\Delta+\omega]_\omega = [\Delta]_\omega + \omega \,,
    \qquad
    [-\Delta]_\omega = -[\Delta]_\omega - 1\,,
\end{equation}
the constraint becomes
\begin{equation}
    [\Delta_1]_\omega + 
    [\Delta_2]_\omega + 
    [\Delta_3]_\omega =     
    2 \omega - \frac{3-n_0}{2}\,.
\end{equation}
In terms of these new chemical potentials the function $\Psi_i$ becomes
\begin{equation}
    \Psi_i = 
    \frac{2 \pi \mi}{\omega} N
    \left([\Delta_2]_\omega + m\right)
    \left([\Delta_3]_\omega + m\right)
    + \frac{2 \pi \mi}{\omega} N m \left(
    \sum_{i \ne j = 1}^N \frac{\hat{u}_{ij}}{N-1} - \frac{\omega}{2}\right)
    ,
    \quad
    m = \frac{1-n_0}{2}\,.
    \label{eq:aiuto}
\end{equation}
Here one could think that, in case II there is an extra term, suggesting a different behavior of the D3-brane backreaction. 
However, once we evaluate such extra term on the basic solution \eqref{basic}, and we sum over $j$, it is irrelevant. This makes the symmetry between the two cases become manifest. 
This result is the BA counterpart of what we obtained by computing the subleading term in the Cardy-like limit. 

We conclude that 
\begin{equation}
    \left.\mathcal{Z}_{2d}\right|_{\text{basic}} =
    \exp \left(
    \frac{2 \pi \mi}{\omega} N 
    \prod_{a=2}^{3} \left( \, [\Delta_a]_{\omega}+m \, \right)
    + \order{N^0}
    \right).
\end{equation}
In order to compare the result with the $4d$ term \eqref{symbasicind}, we rewrite $[\,\cdot\,]_\omega$ in terms of $\{\,\cdot\,\}_\omega$\, as
\begin{equation}
    \left[\Delta\right]_\omega =  \left\{\Delta\right\}_\omega  - 1 \,,
\end{equation}
the constraint \eqref{eq:aiuto} reduces again to \eqref{constr} and the $2d$ index contribution becomes
\begin{equation}
    \left.\mathcal{Z}_{2d}\right|_{\text{basic}} =
    \exp \left(
    \frac{2 \pi \mi}{\omega} N 
    \prod_{a=2}^{3} \left( \, \left\{\Delta_a\right\}_{\omega}-n \, \right)
    + \order{N^0}
    \right),
    \quad
    n = \frac{1+n_0}{2}\,.
\end{equation}
By combining this result with the $4d$ term \eqref{symbasicind}, we write our final result for the defect superconformal index
\begin{equation}
    \left.\log \mathcal{I}\right|_{\text{basic}} = 
    -\frac{\pi \mi}{\omega^2} \, N^2 \,
    \prod_{a=1}^{3} \left(
    \, \left\{\Delta_a\right\}_{\omega}-n \, 
    \right)
    +
    \frac{2 \pi \mi}{\omega} N 
    \prod_{a=2}^{3} \left( \, \left\{\Delta_a\right\}_{\omega}-n \, \right)
    + \log N 
    + \order{N^0}
    \;.
\end{equation}
This result is in agreement with the one obtained in \eqref{eq:finalcardy} in the limit of small $\omega$.


\section{Conclusions}
In this paper we have studied a setup corresponding to a $5d$ rotating BH in presence of surface defects with maximal supersymmetry.
From the holographic perspective the system corresponds to a stack of D3-branes  in AdS$_5 \times S^5$  type IIB supergravity with the addition of a probe D3, extending across both time and the radial direction, while being wrapped around one compact direction in AdS$_5$ and another on the five-sphere.
We have evaluated the SCI of the corresponding dual field theory, consisting in a $4d$-$2d$ system, namely $\SU{N}$ $\mathcal{N}=4$ SYM coupled to a maximal Gukov-Witten surface defect.
We have used two distinct methodologies to evaluated the index: firstly, by considering the Cardy-like limit, and secondly by applying the BA approach matching the two results at large-$N$ for equal angualar momenta. 
Such regimes correspond to the large $N$ limit for equal and small angular velocities. In this case we have extracted the sub-leading logarithmic corrections to the index, that are expected to capture the leading order effect of the backreaction of the probe D3-brane in the dual gravitational picture.  
Furthermore, a three-dimensional EFT emerges from the calculation: the effective picture corresponds to a sum over the  anomalies of the $4d$ and $2d$ system in addition to a pure $\SU{N}_{\pm N}$ topological theory.

There are many open questions and lines of research left. It would be interesting to study the coupling with other GW operators.  So far only the coupling of the four-dimensional theory with a maximal GW operator has been extensively studied. Nevertheless, the possibility of coupling $\SU{N}$ $\mathcal{N}=4$ SYM to GW surface defects corresponding to other Levi sub-gropus 
has been discussed in \cite{Chen:2023lzq}. 
It would be interesting to have an explicit analysis of such Levi subgroups for the GW defects and to understand how this is realized at the level of the SCI by working out the localization procedure of the coupled  $4d$-$2d$ system along the lines of what we have done here.

Another possible extension consists in working out the maximally supersymmetric case for real gauge groups and their connection with the S-duality orbits of SYM.
Although the defect is generally defined by prescribing boundary conditions for the vector field, a useful approach to include such in the SCI is to consider it as a coupling of a $2d$ theory to the four-dimensional one. The standard prescription used here for the  $\mathrm{SU}(N)$ case
could be extended by considering other Lie algebras corresponding to $\mathrm{USp}(2N)$ and $\mathrm{SO}(N)$ gauge groups. 
From the Cardy-like limit of the 4d/2d system SCI it would  require to study then the saddles, similarly to what was done in \cite{Amariti:2023rci} for the pure 4d system.
It would also be interesting to study the fate of S-duality for the $\mathrm{USp}(2N)$ and $\mathrm{SO}(2N+1)$ gauge groups in the coupled system. 

Another generalization of the analysis consists in understanding the behavior of the coupled system around different holonomy saddles than the one treated here. Such saddles admit a holographic interpretation in terms of wrapped D3-brane solutions \cite{Aharony:2021zkr}. Futhermore, an EFT interpretation in terms of orbifold partition functions has been discussed in \cite{ArabiArdehali:2021nsx}. One may wonder the fate of the $4d$-$2d$ coupled system and the role played by the circle reduction of the defect in this case.

While on the Cardy side we have obtained the result for different angular momenta, on the Bethe side we restricted to the case of $\sigma=\tau$. It would be important to extend the BA analysis to the case of $\sigma\neq \tau$. 
Indeed, the BA approach is well-defined also in this more general setup \cite{Benini:2018mlo} and, when we add a defect, the analysis of sub-section \ref{ssbadef} can be generalized, providing a BA formula for the defect SCI.
However, the case of different angular momenta is still an open problem even in absence of defects \cite{Benini:2020gjh,Aharony:2024ntg}.
In fact, the contribution of the basic solution itself requires the evaluation of extra terms that are quite hard to compute.
In \cite{Benini:2020gjh} some of these terms are obtained, and they do not alter the result at leading order.
The recent analysis of \cite{Aharony:2024ntg} has revealed that some other terms are $ \order{N^2}$, thus they cannot be discarded in a large-$N$ limit.
Moreover, the cancellation among the extra terms is argued by focusing on the $\SU{2}$ case.
However, even if these quantities become relevant for large $N$, they are always negligible for large angular momenta.
Therefore, restricting to the comparison with the Cardy-like approach, one could estimate these terms in a double limit of large $N$ and large angular momenta, instead of computing them explicitly. 
Such terms are then discarded, upon ensuring that their recombination is negligible at leading order in the evaluation of the index. 
Following this strategy one could generalize, at least in this double limit, the result obtained here for the defect SCI to the case of different angular momenta.

A last line of research consists in expanding on the three-dimensional EFT interpretation arising from the circle reduction of the parent four-dimensional theory. Our analysis suggests the emergence of an $N$-wound anti-fundamental Wilson loop from the defect. We expect that a complete analysis of the backreaction of the probe D3 will require also an explicit construction of an effective $3d$-$1d$ system.

\section*{Acknowledgments}
We are grateful to Alessia Segati for discussions. The work of the authors has been supported in part by the Italian Ministero dell’Istruzione, Università e Ricerca (MIUR), in part by Istituto Nazionale di Fisica Nucleare (INFN) through the “Gauge Theories, Strings, Supergravity” (GSS) research project.

\appendix
\section{Special functions and asymptotics expansions}
\label{sec:ell_fun}
In this appendix we list general properties and the asymptotic expansions of the special functions used in this work. \\
The $q$-Pochhammer symbol is defined for complex $z,q$ with $|q| < 1$ by
\begin{equation}
\label{Poch}
    (z;q)_\infty \coloneqq \prod_{j = 0} ^ {\infty} \left( 1 - z q^j \right).
\end{equation}
We can derive an asymptotic expansion for the $q$-Pochhammer symbol $(q;q)_\infty$ by rewriting it in terms of the Dedekind Eta function
\begin{equation}
    \eta(\tau) \coloneqq \ee^{\frac{\pi\mi \tau}{12}}\prod_{n = 1}^{\infty}(1 - \ee^{2n\mi\pi\tau}),
\end{equation}
and employing its modular properties
\begin{equation}
    (q;q)_\infty = \ee^{-\frac{ \pi \mi \tau}{12}}\eta(\tau) \underset{r \to 0}{\sim} - \frac{\pi \mi}{12}\left(\tau + \frac{1}{\tau}\right) - \frac{1}{2}\log(-\mi\tau).
\end{equation}\\
Similarly $\theta_0(u;\tau)$ is defined as
\begin{equation}
    \theta_0(u;\tau) \coloneqq 
    (\ee^{2 \pi \mi u};\ee^{2 \pi \mi \tau})_{\infty} \,
    (\ee^{2 \pi \mi \tau} \ee^{-2 \pi \mi u};\ee^{2 \pi \mi \omega})_{\infty}.
\end{equation}
It satisfies the quasi-double periodicity property
\begin{equation}
    \label{eq:thetaper}
    \theta_0(u + m + n\tau;\tau)=(-1)^n\ee^{-2\pi\mi n u}\ee^{-\pi\mi n (n-1)\tau}\theta_0(u;\tau), \quad m,n \in \mathbb{Z}
\end{equation}
and the inversion formula
\begin{equation}
    \label{eq:thetainv}
    \theta_0(-u;\tau) = -\ee^{2\pi\mi u}\theta_0(u;\tau).
\end{equation} \\
In addition, we remind the relation between $\theta_0(u;\tau)$ and the Jacobi theta function $\theta_1(u;\tau)$:
\begin{equation}
    \label{eq:theta01}
    \theta_1(u;\tau) = \mi \ee^{\pi\mi \tau/4 - \pi\mi u}(q;q)_\infty \theta_0(u;\tau).
\end{equation}
The elliptic gamma function is defined as
\begin{equation}
    \Gamma(z;p,q) \coloneqq 
    \prod_{m=0}^{\infty}\prod_{n=0}^{\infty}
    \frac{ 1 - p^{m+1}q^{n+1}/z }{ 1 - p^{m}q^{n}z } \, ,
    \qquad 
    \widetilde{\Gamma}(u) \coloneqq \Gamma(\ee^{2 \pi \mi u}; \ee^{2 \pi \mi \tau}, \ee^{2 \pi \mi \sigma}).
    \label{gamma}
\end{equation}
Similarly to $\theta_0(u,\tau)$, also elliptic gamma function satisfies an inversion formula
\begin{equation}
    \label{eq:gammainv}
    \widetilde{\Gamma}(u;\tau,\sigma) =  \widetilde{\Gamma}(\sigma + \tau - u;\tau,\sigma)^{-1},
\end{equation}
and a quasi double-periodicity relation
\begin{equation}
    \label{eq:gammaper}
    \widetilde{\Gamma}(u;\tau,\sigma) = \theta_0(u;\tau)^{-1}\widetilde{\Gamma}(u + \sigma;\tau,\sigma) = \theta_0(u;\sigma)^{-1}\widetilde{\Gamma}(u + \tau;\tau,\sigma).
\end{equation}
Using \eqref{eq:gammainv} and \eqref{eq:gammaper} together with \eqref{eq:thetaper} and \eqref{eq:thetainv} one obtains
\begin{equation}
    \label{pmad}
    \sum_{i\neq j}\log\widetilde{\Gamma}(u_{ij};\tau,\sigma) = -\sum_{i < j}\left(\log\theta_0(u_{ij};\tau)+\log\theta_0(-u_{ij};\sigma)\right).
\end{equation}
Exploiting the modular properties of $\theta_0(u;\tau)$ one derives the asymptotic expansion for small $\tau$
\begin{eqnarray}
    \log\theta_0(u;\tau) =&& \frac{\mi \pi }{\tau} \{u\}_\tau (1-\{u\}_\tau) + \mi \pi \{u\}_\tau-\frac{\mi \pi }{6\tau} \left(1 + 3\tau + \tau^2\right) + \nonumber \\
    && + \log \left( \left(1 - \ee ^ { - \frac{ 2\pi \mi }{\tau} \{u\}_\tau} \right) \left(1 - \ee ^{ - \frac{2\pi \mi}{\tau}(1 - \{u\}_\tau)} \right) \right) + \mathcal{O} \left(\ee^{ - \frac{2 \pi \sin{ \arg(\tau)} }{|\tau|}} \right),\nonumber\\
\end{eqnarray}
where
\begin{equation}
    \{u\}_\tau \equiv \{\tilde{u}\} + \tau \bar{u}, \quad u \equiv \tilde{u} + \tau \bar{u}, \quad \tilde{u},\bar{u}\in\mathbb{R}
\end{equation}
and $\{\tilde{u}\} = \tilde{u} - \lfloor u \rfloor$.
For small $\tau\ne\sigma$, such definition is generalized to 
\begin{equation}
    \{x\} = \{\tilde x\} + r\bar x\equiv \tilde x - \lfloor \tilde x \rfloor + r \bar x
\end{equation}
for any $x$ with $\tilde x \neq 0$.
To recover an asymptotic expansion for the elliptic gamma function one can start from the infinite product formula 
\begin{equation}
    \widetilde{\Gamma}(u;\tau,\sigma)=
    \ee^{2 \pi \mi Q(u;\tau,\sigma)}
    \prod_{n=-\infty}^{\infty}
    \ee^{
        -\text{sign}(n)
        \frac{\pi \mi}{2 \tau \sigma}
        \left(
            \left(
                \frac{u+n}{r}-\frac{\tau+\sigma}{2}
            \right)^2-
            \frac{\tau^2+\sigma^2}{12}
        \right)
    }
    \Gamma_h\left(
        \frac{u+n}{r};\omega_1,\omega_2
    \right),
\end{equation}
where $\Gamma_h(u;\omega_1,\omega_2)$ is the hyperbolic gamma function. As $r$ approaches zero the infinite tower of KK modes associated with the hyperbolic gamma functions gets lifted, when $u \not\in \mathbb{Z}$, and we get
\begin{equation}
\log\widetilde\Gamma(u;\tau,\sigma) = 2\pi \mi Q(\{u\}) + \mathcal{O}(\ee^{-1/r})
\end{equation}
with
\begin{eqnarray}
    Q(\{u\}) =&& -\frac{B_3(\{u\})}{6 \,\sigma\tau } \,+\, B_2(\{u\}) \frac{(\sigma + \tau )}{4\, \sigma\tau } \,-\, B_1(\{u\})\frac{ \left((\sigma + \tau )^2 + \sigma \tau \right)}{12\, \sigma  \tau } \nonumber \\ 
    && +\, \frac{\sigma }{24} \,+\, \frac{\tau }{24},
\end{eqnarray}
and the Bernoulli polynomials 
\begin{equation}
    B_3(u) = u^3 - \frac{3}{2}u^2 + \frac{u}{2}, \quad B_2(u) = u^2 - u + \frac{1}{6}, \quad B_1(u) = u - \frac{1}{2}.
\end{equation}
\section{Wilson loop in pure CS}
\label{sec:wils_loop}
In this appendix we want to evaluate the partition function of a $SU(N)_{n_0 N}$ pure CS theory with an (anti-)fundamental $N$-wounded Wilson loop insertion, where $n_0 = \pm 1$. We start by first recalling the result for the three-sphere partition function of a pure $3d$ CS theory.
The squashed three-sphere partition function of a pure $\U{N}_{n_0 N}$ CS theory is 
\begin{equation}
    \label{eq:partUN}
    \frac{1}{N!} 
    \int \prod_{i=1}^{N} \frac{\differential \lambda_i}{\sqrt{- \omega_1\omega_2}} \frac{\ee^{- \frac{\pi \mi n_0 N}{\omega_1\omega_2} \sum_{i=1}^{N}\lambda_i^2}}{\prod_{i < j}  \Gamma_h( \lambda_{ij})\Gamma_h( - \lambda_{ij})}.
\end{equation}
We can constrain the holonomies with a Lagrange multiplier to derive the partition function for the case of $\SU{N}$ gauge group:
\begin{equation}
    \label{eq:partSUN}
    \frac{1}{N!} 
    \int \dd \Lambda \int \prod_{i=1}^{N} \frac{\differential \lambda_i}{\sqrt{- \omega_1\omega_2}} \frac{\ee^{- \frac{\pi \mi n_0 N}{\omega_1\omega_2} \sum_{i=1}^{N}\lambda_i^2 + 2\pi \mi \Lambda \sum_{j = 1}^{N}\lambda_j}}{\prod_{i < j}  \Gamma_h( \lambda_{ij})\Gamma_h( - \lambda_{ij})}.
\end{equation}
Employing the identity
\begin{equation}
  \frac{1}{\Gamma_h(x)\Gamma_h( -x)} = -4\sin(\frac{\pi x}{\omega_1})\sin(\frac{\pi x}{\omega_2})
\end{equation}
and specializing to the case $\omega_1 = \omega_2$ we get
\begin{equation}
    Z_{\SU{N}_{n_0 N}}^{S^3} =\ee^{ \frac{5\pi i n_0(N^2 - 1)}{12}}.
\end{equation}

Let us introduce a Wilson loop operator in the CS theory.
The supersymmetric Wilson loop is defined as 
\begin{equation}
    W_\gamma(\sigma) = \Tr_R P \exp{\oint_\gamma i A_\mu \dot{x}^{\mu} + \sigma |\dot{x}|\dd \tau }.
\end{equation}
The localization locus for a gauge theory on $S^3$ is defined by the equations
\begin{equation}
    \begin{cases}
        F_{\mu\nu} = 0\\
        D = -\sigma \equiv -\sigma_0.\\
    \end{cases}
\end{equation}
Thus, an n-wounded Wilson loop insertion in the functional integral modifies the matrix model arising from localization with a term 
\begin{equation}
    W_\gamma(\sigma_0) = \Tr_R \exp{\sigma_0\oint\dd s} =  \Tr_R \exp{2\pi n \sigma_0}.
\end{equation}
Let us compute the partition function of a k-level CS-theory with an n-wounded Wilson loop in the (anti-)fundamental representation of $\U{N}$. Starting from \eqref{eq:partUN} and using the Weyl denominator formula
\begin{equation}
    \prod_{1\leq i < j \leq N} 2 \sinh \left(\frac{x_i - x_j}{2}\right) = \sum_{\sigma}(-1)^\sigma\prod_{j}\ee^{((N+1)/2 - \sigma(j))x_j},
\end{equation}
where the sum runs over the permutations $S_N$, we get
\begin{equation}
    \begin{split}
        Z_W = \frac{1}{N!}\int & \prod_{j=1}^{N} \dd \lambda_j e ^ {-i \pi k \lambda_j^2} \sum_{\sigma_1, \sigma_2}(-1)^{\varepsilon(\sigma_1) + \varepsilon(\sigma_2)} \\
        &\prod_{j=1}^{N} \ee^{2\pi(N + 1 - j -\sigma_1(j) - \sigma_2(j))\lambda_j}
        \left(\sum_{i=1}^{N} \ee^{2\pi n \lambda_i}\right)
    \end{split}
\end{equation}

Since $ \left(\sum_{i=1}^{N} \ee^{2\pi n \lambda_i}\right)$ is symmetric under exchange of $\lambda_i$ with $\lambda_j$ we can freely relabel variables as before to get rid of one sum over permutations, without spoiling the result.

We get
    \begin{equation}
        \begin{split}
            Z_{W}^{\U{N}} = &\sum_{i=1}^N \sum_{\sigma\in S_N}(-1)^{\varepsilon(\sigma)} \int \left(\dd{\lambda_i}\ee^{-ik\pi \lambda_i^2}\ee^{2\pi(N+1-i-\sigma(i))\lambda_i}\right)\\
            &\int \left(\prod_{j\neq i}^N\dd{\lambda_j}\ee^{-ik\pi \lambda_j^2}\ee^{2\pi(N+1-j-\sigma(j))\lambda_j}\right) = \\
            =& (i k)^{-N/2}\sum_{i=1}^N \sum_{\sigma\in S_N}(-1)^{\varepsilon(\sigma)}\prod_{j=1}^N \ee^{-\frac{i\pi}{k}\left(N + 1 + n\delta_{i,j} - j -\sigma(j)\right)^2}
        \end{split}
    \end{equation}
Expanding the square
\begin{equation}
    \sum_{j=1}^N (N + 1 - j + n\delta_i,j - \sigma(j))^2 =  \sum(x_j^2 - 2x_j\sigma(j) + j^2),
\end{equation}
where $x_j = N + 1 - j + n\delta_{i,j}$, we isolate a term independent on $\sigma(j)$ and a term $x_j\sigma(j)$, which can be rearranged with the Weyl denominator formula after being combined with the sum over $\sigma$.

We get 
\begin{equation}
    \begin{split}
        Z_{W}^{\U{N}} =& (ik)^{-N/2}\sum_{i=1}^N \ee^{-\frac{i\pi}{k}\sum_j (x_j^2 + j^2)} \sum_{\sigma\in S_N}(-1)^{\varepsilon(\sigma)} \ee^{\frac{2\pi i }{k}\sum_j x_j \sigma(j)} = \\
        =& (ik)^{-N/2}\sum_{i=1}^N \ee^{-\frac{\pi i}{k}\sum_j (x_j^2 + j^2)- (N + 1) x_j} \prod_{i<j} 2\sinh\left(\frac{x_i - x_j}{2}\right) = \\
        =& (ik)^{-N/2} \ee^{-\frac{\pi i}{6k}\left(N(N^2 - 1)+ 6n(N + 1 + n) \right)}(-i)^{N(N-1)/2}\\
        &\sum_{j=1}^N \ee^{ \frac{2 \pi i}{k} j n}\prod_{m<l} 2\sin\left(\frac{\pi}{k}(l - m + n (\delta_{j,m} - \delta_{j,l})) \right)
    \end{split}
\end{equation}
\begin{equation}
    \begin{split}
       & \prod_{m<l} 2\sin\left(\frac{\pi}{k}(l - m + n (\delta_{j,m} - \delta_{j,l})) \right) =\\
       &=\prod_{m<l,\,m,l\neq j} 2\sin\left(\frac{\pi}{k}(l - m ) \right)\prod_{m=1}^{j-1} 2\sin\left(\frac{\pi}{k}(j - m - n) \right) \prod_{l=j+1}^{N} 2\sin\left(\frac{\pi}{k}(l - j + n) \right) =\\
       &=\prod_{m<l} 2\sin\left(\frac{\pi}{k}(l - m ) \right)\prod_{m=1}^{j-1}\frac{ \sin\left(\frac{\pi}{k}(j - m - n) \right)}{ \sin\left(\frac{\pi}{k}(j - m) \right)} \prod_{m=j+1}^{N} \frac{\sin\left(\frac{\pi}{k}(m - j + n) \right)}{\sin\left(\frac{\pi}{k}(m - j) \right)} = \\
       &= \prod_{m<l} 2\sin\left(\frac{\pi}{k}(l - m ) \right)\prod_{m\neq j}^{N}\frac{ \sin\left(\frac{\pi}{k}(m - j + n) \right)}{ \sin\left(\frac{\pi}{k}(m - j) \right)}
    \end{split}
\end{equation}
Then, the final result is
\begin{equation}
    \label{eq:Z_W}
    \begin{split}
    Z_{W_n}^{\U{N}_k} = & (ik)^{-N/2} \ee^{-\frac{\pi i}{6k}\left(N(N^2 - 1)+ 6n(N + 1 + n) \right)}(-i)^{N(N-1)/2}\\
    &\prod_{m<l} 2\sin\left(\frac{\pi}{k}(l - m ) \right) \sum_{j=1}^N \ee^{ \frac{2 \pi i}{k} j n} \prod_{m\neq j}^{N}\frac{ \sin\left(\frac{\pi}{k}(m - j + n) \right)}{ \sin\left(\frac{\pi}{k}(m - j) \right)}
    \end{split}
\end{equation}
Let us consider the gapped case for the pure CS theory $k = N$ (the case $k = -N$ goes along the same line). Then, (\ref{eq:Z_W}) is zero unless the wounding of the Wilson loop is $n = p N $,
\begin{equation}
    \begin{split}
    Z_{W_{pN}}^{\U{N}_N} = & (N)^{-N/2} \ee^{-\frac{\pi i}{6}\left((N^2 - 1)+ 6p(N(p+1) + 1) \right)}\ee^{-i\pi N^2/4}\\
    & (-1)^{N-1} N \prod_{m<l} 2\sin\left(\frac{\pi}{N}(l - m ) \right)
    \end{split}
\end{equation}
Notice that $\prod_{m<l} 2\sin\left(\frac{\pi}{N}(l - m )\right) = N^{N/2}$ and we get
\begin{equation}
    Z_{W_{pN}}^{\U{N}_N} = (-1)^{N-1 + p(N(p+1)+1)} N Z_{CS}^{\U{N}}.
\end{equation}
When $p = \pm 1$ we have
\begin{equation}
    Z_{W_{\pm N}}^{\U{N}_N} = (-1)^{N} N Z_{CS}^{\U{N}_N}.
\end{equation}
Ultimately we specialize to $\SU{N}$ by introducing the Lagrange multiplier $\Lambda$ as in \eqref{eq:partSUN}, and we obtain
\begin{equation}
    \begin{split}
        Z_{W_n}^{\SU{N}_k} = Z_{W_n}^{\U{N}_k} \ee^{-\frac{i \pi  n^2}{k N}}\sqrt{\frac{i k}{N}}.
    \end{split}
\end{equation}
We are interested in the case $n = -N$ and $k = N$:
\begin{equation}
    Z_{W_{-N}}^{\SU{N}_N} = (-1)^{N + 1} N Z_{CS}^{\SU{N}_N}.
\end{equation}
\bibliographystyle{JHEP}
\bibliography{ref.bib}
\end{document}